\documentclass[structabstract]{aa}

\usepackage{graphicx}
\usepackage{rotating}
\usepackage{enumerate}
\usepackage[english]{babel}
\usepackage[T1]{fontenc}
\usepackage{txfonts}
\usepackage{natbib}
\bibpunct{(}{)}{;}{a}{}{,}
\begin{document}
%

%\titlerunning{Suzaky observations of Mrk 509 and Mrk 841}
   \title{Suzaku broad-band observations of the Seyfert 1\\ galaxies Mrk~509 and Mrk~841}

   \author{M. Cerruti
          \inst{1}
          \and
          G. Ponti\inst{2}
          \and
          C. Boisson\inst{1}
          \and
          E. Costantini\inst{3}
          \and
          A.L. Longinotti\inst{4}
          \and
          G. Matt\inst{5}
          \and
          M. Mouchet\inst{6,1}
          \and
          P.O. Petrucci\inst{7}
          }

   \institute{LUTH, Observatoire de Paris, CNRS, Université Paris Diderot; 5 Place
Jules Janssen, 92190 Meudon, France
\\
              \email{matteo.cerruti@obspm.fr}
         \and
   			     School of Physics and Astronomy, University of Southampton, Highfield, Southampton SO17 1BJ, UK
         \and
             SRON, Netherlands Institute for Space Research, Sorbonnelaan 2, 3584 CA Utrecht, The Netherlands
         \and
             MIT, Kavli Institute for Astrophysics and Space Research, 77 Massachusetts Avenue, NE80-6011, Cambridge, MA, 02139, U.S.A.
         \and
             Dipartimento di Fisica, Università degli Studi Roma Tre, Via della Vasca Navale 84, 00146 Roma, Italy
         \and             
             APC, Université de Paris 7 Denis Diderot, 75205 Paris Cedex, France
   			 \and
   			     UJF-Grenoble 1 / CNRS-INSU, Institut de Planétologie et d'Astrophysique de Grenoble (IPAG) UMR 5274, Grenoble, F-38041, France	
             }   
             
 \date{Received 6 January 2011/
       Accepted 22 August 2011 }
 
  \abstract
   {Markarian $509$ and Markarian $841$ are two bright Seyfert 1 galaxies with X-ray spectra characterised by a strong soft excess and a variable Fe $K\alpha$ line, as shown by several X-ray observatories in the past.}
   {We report an analysis and modelling of new \textit{Suzaku} observations of these sources, taken between April and November, 2006, for Mrk~509, and between January and July, 2007, for Mrk~841, for a total exposure time of $\approx 100$ kiloseconds each. Data from \textit{XIS} and \textit{HXD/PIN} instruments, going from $0.5$ to $60$ keV, represent the highest spectral resolution simultaneous broad-band X-ray spectrum for these objects, and provide the strongest constraints yet on the origin of the soft excess emission.}
   {We fitted the broad-band spectrum of both sources with a double Comptonisation model, adding neutral reflection from distant material and a two-phase warm absorber. We then studied the two competing models developed to explain the soft excess in terms of atomic processes: a blurred ionised disc reflection and an ionised absorption by a high velocity material.} 
   {When fitting the data in the $3$-$10$ keV range with a power law spectrum, and extrapolating this result to low energies, a soft excess is clearly observed below $2$ keV, although its strength is weak compared to previous observations of both sources. A moderate hard excess is seen at energies higher than $10$ keV, together with a neutral Fe $K\alpha$ narrow emission line at $E_0 \approx 6.4$ keV and a broad Fe emission line. For Mrk~509, the broad Fe emission line is required in all the three physical models to ensure a good fit to the data: this finding suggests that the blurred reflection model correctly describes the soft excess, but that it underestimates the broad Fe emission line. For the smeared absorption model, this suggests instead that the continuum spectrum absorbed by the outflowing gas should indeed contain a reflected component. For Mrk~841, all three models that we tested provide a good fit to the data, and we cannot rule out any of them. A broad emission line is required in the double Comptonisation and smeared absorption models, while the blurred reflection model consistently fits the broad-band spectrum, without adding any extra emission-line component.}
  {}

   \maketitle
  
   \section{Introduction}
   The origin of the soft X-ray excess emission is still nowadays, about three decades after the discovery \citep{Pravdo,Arnaud85}, one of the major open questions in AGN (active galactic nuclei) research \citep[e.g.][]{Turner09}. This component appears as a featureless excess of emission above the low energy extrapolation of the $2$-$10$ keV best-fit power law. Historically associated with the high energy tail of the accretion-disc black-body radiation, many authors (\citealp{Czerny03}; \citealp[hereafter GD04]{Gierlinski04}; \citealp{Piconcelli05}; \citealp[hereafter C06]{Crummy06}) have shown that modelling with a thermal continuum infers a characteristic ''temperature'' higher than expectations \citep{Shakura76} that remains remarkably constant across a range of AGN despite a wide spread in black hole mass and AGN luminosity. Moreover, in bright and variable AGN the soft excess does not follow the expected black-body luminosity-temperature relation \citep{Ponti06}. Various studies have shown that the ratio of the soft excess, at $0.5$ keV, to the extrapolation of the high-energy power-law emission has a very small scatter \citep{Piconcelli05, Miniutti09}, which is very different from what is observed in galactic black holes in the bright soft state, dominated by the disc black-body emission \citep[e.g.][]{DoneCYG}. These findings triggered the search for a different origin of the soft excess. The first idea investigated was the Comptonisation that might occur in the upper layer of the accretion disc (\citealp{Czerny03}; GD04; \citealp{Sobolewska07}). This model may explain the featureless shape of the soft excess and its high temperature; the feedback between the coronal disc irradiation and the disc skin may provide a mechanism to stabilize both the soft excess temperature and the ratio relative to the power-law emission. On the other hand, the observed constancy of the soft excess temperature seems to suggest a nature tied to atomic processes. Two possible alternatives are reflection and absorption. If the upper layer of the accretion disc is ionized, the disc Compton reflection component will contain many X-ray lines \citep{RossFabian} that will be broadened by the relativistic motion of the material on the surface of the accretion disc, forming a featureless continuum (C06). Ionized absorption processes can also imprint their main features in the soft X-ray band. If these absorbers are moving relativistically (for example, in the form of a disc wind) they may, as well, produce a featureless curved continuum and reproduce the soft excess. Nevertheless, numerical simulations indicate that, to reproduce the excess, the velocities of such a wind have to be extremely high \citep{Schurch08}, suggesting that the flow has to be clumpy and/or only partially covering the source, or associated with a magnetically driven jet.\\
   
   These different models are degenerate in the $0.5$-$10$ keV band but still predict different amounts of hard X-ray emission. In addition, they predict different variability behaviours in the soft and hard energy band and, thus, the measurement of the high energy flux may help us to determine the physical origin of the soft excess. To this aim, with \textit{Suzaku} we observed Markarian 509 (Mrk~509), the brightest Seyfert 1 galaxy in the hard X-ray sky ($17$-$60$ keV) with an important soft excess but no strong warm absorber component \citep{Sazonov07}, and Markarian 841 (Mrk~841), one of the first AGN in which a soft excess has been discovered \citep{Arnaud85}. Information about the four Mrk~509 observations and the two observations of Mrk~841 are given in Table \ref{observationlog}.\\
      \begin{table}
   \caption{\textit{Suzaku} observation log}              % title of Table
   \label{observationlog}      % is used to refer this table in the text
   \begin{tabular}{c c c c c}
   \hline
   \multicolumn{5}{c}{Mrk~509}\\
   \hline
   & ID & Aim Point & Start &  Exposure (s) \\
   \hline
   obs.1 & $701093010$ & HXD & $2006$-$04$-$25$ &  $24580$ \\
   obs.2 & $701093020$ & XIS & $2006$-$10$-$14$ &  $25960$ \\
   obs.3 & $701093030$ & XIS & $2006$-$11$-$15$ &  $24450$ \\
   obs.4 & $701093040$ & XIS & $2006$-$11$-$27$ &  $33090$ \\
   \hline
   tot   &             &     &                  &  $108080$ \\
   \hline
   \hline
   \multicolumn{5}{c}{Mrk~841}\\
   \hline
   & ID & Aim Point & Start &  Exposure (s) \\
   \hline
   obs.1 & $701084010$ & HXD & $2007$-$01$-$22$ &  $51790$ \\ 
   obs.2 & $701084020$ & HXD & $2007$-$07$-$23$ &  $50930$ \\
   \hline
   tot   &             &     &                  &  $102720$ \\
   \hline
   \end{tabular} 
   \end{table}
   
   Located at a redshift of $z = 0.0344$ \citep{Fisher95}, Mrk~509 was first observed at X-ray energies by \textit{Ariel V} \citep{Cooke78}, and then studied by several X-rays telescopes. Using \textit{HEAO-1}, \citet{Singh85} detected its soft excess for the first time, a detection that was later confirmed by \textit{EXOSAT} \citep{Morini87}, in addition to the detection of a Fe $K\alpha$ emission line. Using \textit{Ginga}, \citet{Singh90} revealed a hardening of the power law at energies greater than $10$ keV. \textit{BeppoSAX} observations \citep{Perola00} showed a power-law cut-off at $70$ keV and revealed information about the warm absorber \citep{Dadina05}, which since has been studied in depth by \citet{Smith07} and \citet{Detmers10}, with \textit{XMM-Newton/RGS} data. \citet{Ponti09} presented the most complete study to date of the Fe K complex, using \textit{XMM-Newton} data and the same \textit{Suzaku} observations we present here. For this reason, we focus on the broad-band spectrum only, and no further analysis of emission lines will be done. Recently, \citet{Kaastra11} carried out a multi-wavelength campaign of this source, using data from seven different satellites and observatories (from infra-red to hard X-rays, see \citet{Kaastra11} and references therein).\\
   
The soft excess of the bright Seyfert 1 galaxy Mrk~841 ($z = 0.0365$) was first detected by \textit{EXOSAT} \citep{Arnaud85}. As for Mrk~509, Mrk~841 has been extensively studied. \citet{George93} provided the first evidence of Fe $K\alpha$ emission-line detection and broad-band variability. In a sample of 24 type~1 AGN observed by \textit{ASCA} \citep{Reynolds97}, Mrk~841 belongs to sources for which a ''statistically significant soft excess'' is observed. More recent observations have been performed by \textit{BeppoSAX} \citep{Bianchi01} and \textit{XMM-Newton} (\citealp[hereafter P07]{Petrucci07}; \citealp{Longinotti10}){:} P07 studied in detail the Fe K complex variability, while Longinotti et al., using \textit{XMM-Newton/RGS} data, investigated the warm absorber structure. Even though the goal of this paper is to investigate the broad-band X-ray spectrum of this object, in Section $3.4$ we also discuss the Fe $K\alpha$ narrow emission-line detection.\\

The paper is organised as follows. In Section 2, we detail the \textit{Suzaku} data reduction and present the serendipitous detection of the blazar PKS 1502+106. In Section 3.1, we study the light curves of Mrk~509 and Mrk~841; in Section 3.2, we analyse the soft X-ray excess in both sources and in Section 3.3 we discuss their hard excess; in Section 3.4, we concentrate on the Fe $K\alpha$ narrow emission line detection in Mrk~841; Section 3.5 is dedicated to the modelling of the broad-band (\textit{XIS} and \textit{PIN}) Suzaku summed spectra in a double Comptonisation scenario, discussing the detection of a warm absorber in both sources; in Sections 3.6 and 3.7, we fit the broad-band summed spectra with a blurred ionised reflection model and a smeared ionised absorption one. The discussion and the conclusions are given in Section 4.\\
  
     \begin{figure}[t!]
	 \centering
   \resizebox{\hsize}{!}{\includegraphics[width=17cm, angle=0]{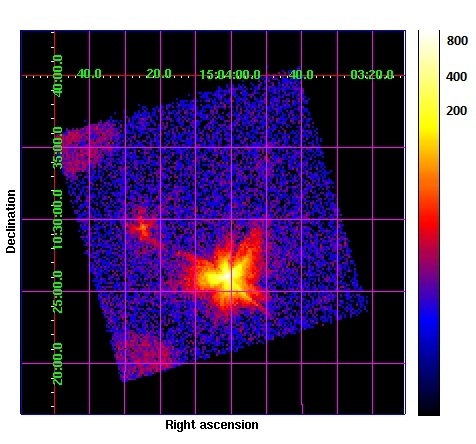}}
     \caption{\textit{Suzaku/XIS0} image of Mrk~841 (first observation): the brightest source is Mrk~841 and the fainter one is consistent with PKS 1502+106; the two calibration spots are visible in the top and bottom left corners. }
     \label{MRK841XIS0}
   \end{figure} 
   
   \section{Data reduction}
    \textit{Suzaku} X-ray telescope carries two main instruments: the \textit{XIS} \citep{XIS}, a soft ($0.2$-$12$ keV) X-ray imaging spectrometer, and the \textit{HXD} \citep{HXD}, a hard X-ray detector that covers the higher energy domain ($10$-$500$ keV).\\  
    
    The \textit{XIS} instrument is composed of four CCD cameras: three of them (\textit{XIS0,2,3}) are front-illuminated (\textit{FI}), while one (\textit{XIS1}) is back-illuminated (\textit{BI}). Owing to an instrumental failure, \textit{XIS2} is unavailable for observations performed after November 9, 2006: for uniformity between observations, we did not consider data from this instrument.
   Event files from version 2.0.6.13 of the \textit{Suzaku} pipeline processing were used, with standard screening criteria, and data were then reduced using version 6.7 of \textit{HEASoft}\footnote{http://heasarc.nasa.gov/lheasoft}. Spectra and light curves were extracted using \textit{XSelect}, version 2.4a. The chosen extracting region is large enough ($r$$>$$150$ arcsec) to avoid calibration problems (as suggested in the \textit{Suzaku} ABC Guide\footnote{http://heasarc.gsfc.nasa.gov/docs/suzaku/analysis/abc/}). The background was extracted from a region (circular or annular, according to the aim point) in the same image, respecting the same geometrical constraints as for the source. Response matrices and ancillary response files were generated for each \textit{XIS} using \textit{xisrmfgen} (version 2009-02-28) and \textit{xissimarfgen} (version 2009-01-08) tools. The two \textit{XIS/FI} spectra were added using \textit{mathpha} (version 4.1.0.) with the \textit{'POISS-3'} error propagation method, which represents a mean value between the two extreme error values \textit{'POISS-1'} and \textit{'POISS-2'} (see \textit{mathpha} user's guide\footnote{http://heasarc.gsfc.nasa.gov/docs/heasarc/caldb/docs/memos/ogip 95 008/ogip 95 008.pdf} and references therein). Response files were added using \textit{addrmf}, version 1.21, and \textit{addarf}, version 1.2.6.  A single \textit{XIS} lightcurve was obtained by summing up the three \textit{XIS0,1} and \textit{3} background-subtracted lightcurves, using \textit{XRonos}, version 5.22.\\
   
   The \textit{HXD} instrument is composed of two detectors: \textit{PIN} (positive intrinsic negative diodes; $10$-$60$ keV) and \textit{GSO/BGO} (gadolinium silicate/bismuth germanate crystals; $50$-$500$ keV). In this paper, we present data from \textit{PIN} only, since no significant signal was detected from \textit{GSO}. For \textit{HXD/PIN}, non-X-ray instrumental background (NXB) and response matrices provided by the \textit{HXD} instrument team were used. An additional component for the CXB (cosmic X-ray background) \citep{CXB} was included and added to the NXB using \textit{mathpha}. Spectra were corrected for the dead time, using \textit{hxddtcor}, version 1.50, and the background exposure time was increased by a factor of ten, following the \textit{Suzaku} ABC Guide. For Mrk~841 observations, PKS 1502+106 (see section 2.1) is in the \textit{PIN} field of view, and its flux, as extrapolated from \textit{XIS} observations, was added with \textit{mathpha} to the background.\\
   
    Each spectrum was rebinned according to the following procedure: for \textit{XIS}, we checked the true energy resolution of each instrument, extracting the spectrum of the calibration spots in the CCD camera (see Fig. \ref{MRK841XIS0}), and measuring the energy and the FWHM (full width at half maximum) of the Mn $K\alpha$ line; with these values, and using equation $(1)$ given in \cite{XIS}, we get the true \textit{XIS} energy resolution as a function of energy; for \textit{PIN}, we used energy resolution values given in \citet{HXD}; we then rebinned the data to have five bins for each energy resolution element. To use a $\chi^2$ minimization algorithm, we imposed a minimum number of counts per bin: this second rebinning does affect the first and the last bins of the spectra only, while in the other channels the condition of five bins per resolution element is always dominant. We chose a minimum of 30 (for Mrk~841) and 50 (for Mrk~509) counts per bin, leading to a similar number of bins for both sources. We considered \textit{XIS/FI} data in the energy band $0.6$-$10$ keV, and \textit{XIS1} data in the range $0.5$-$8.5$ keV; \textit{XIS} data comprised between $1.62$ and $1.82$ keV were ignored owing to a known calibration problem around the Si K edge\footnote{http://heasarc.nasa.gov/docs/suzaku/analysis/sical.html}; \textit{PIN} data below $15$ keV were ignored, and the last \textit{PIN} significant bin is at $\approx$$33$ keV for Mrk~841 and $\approx$$38$ keV for Mrk~509.\\
   
  We also studied the summed spectra of the sources, adding \textit{XIS} and \textit{PIN} data from different observations. For Mrk~509, the first observation was not included in the summed spectrum, because it had been performed using the \textit{HXD} aim point, instead of the \textit{XIS} aim point used for the three other observations (see Table \ref{observationlog}). Source and background spectral files from different observations were added using the same procedure described above for the sum of \textit{XIS/FI} spectra.\\
  To take into account the relative normalization between different instruments during the fitting, models for each instrument data were multiplied by a constant, as indicated by Maeda et al. (2008) \footnote{http://www.astro.isas.ac.jp/suzaku/doc/suzakumemo/suzakumemo-2008-06.pdf}. Spectral fitting was performed with \textit{XSpec}, version 12.5.1. Parameter errors in both text and tables are given at the $90\%$ confidence level.\\
  We assume a flat cosmology with $H_0 = 70\ \textrm{km}\ \textrm{s}^{-1} \textrm{Mpc}^{-1}$ and $\Omega_{\Lambda}=0.73$.\\

   \subsection{Contaminating source: PKS 1502+106}
   
      \begin{figure}[t]
	 \centering
   \resizebox{\hsize}{!}{\includegraphics[width=17cm, angle=-90]{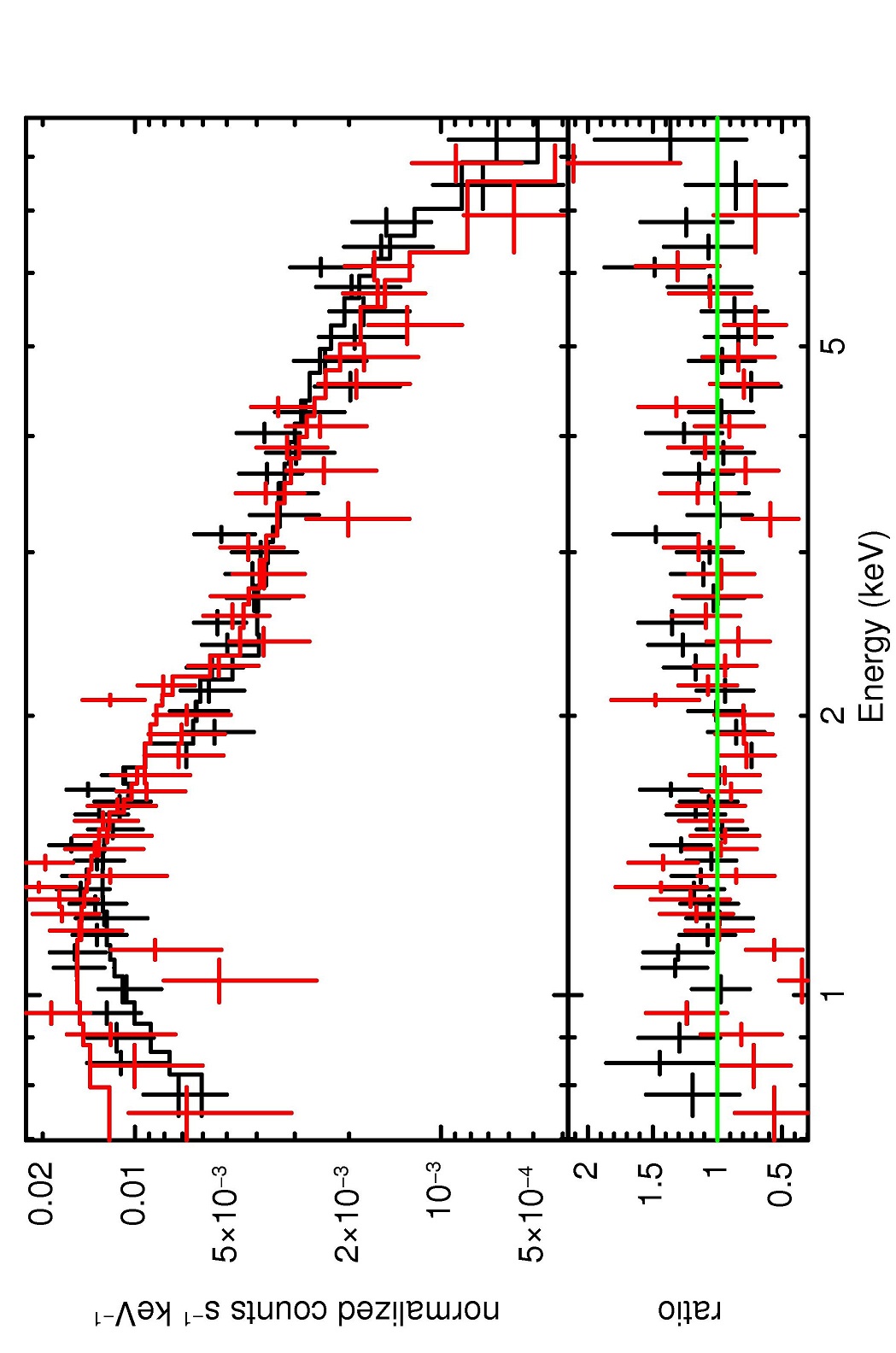}}
     \caption{PKS 1502+106 spectrum (observer frame) as seen by \textit{Suzaku/XIS/FI} (black) and \textit{XIS1} (red). Data were fitted with a power-law model absorbed by galactic material (see section 2.1): data-to-model ratio is shown in the bottom plot.}
     \label{PKS1502_spectrum}
   \end{figure}

   Figure \ref{MRK841XIS0} shows the \textit{Suzaku/XIS0} image of the first observation of Mrk~841. The brightest source at the \textit{HXD} aim point position is Mrk~841, while a second bright source appears in the field of view (at a distance of about $7'$ from Mrk~841), that is coincident with PKS 1502+106. This source, situated at $z=1.839$ \citep{Smith77}, was classified as a blazar by \citet{Abdo10}. We extracted the spectrum of this source following the same procedure discussed in the previous section: Fig. \ref{PKS1502_spectrum} shows the XIS spectrum of PKS 1502+106. We fitted the spectrum with a simple power law absorbed by Galactic material (model \textit{wabs*powerlaw} in \textit{Xspec}, with $N_H = 2.36 \cdot 10^{20}\ \textrm{cm}^{-2}$, \citet{Dickey90}). The best-fit photon index is $\Gamma = 1.32 \pm 0.08$ ($\tilde{\chi^2}$=$\chi^2/\ (\textrm{degrees of freedom (DOF))}$=$71/86$). The source flux in the $0.5$-$2$ keV and $2$-$10$ keV bands is $2.2^{+0.3}_{-0.3}$ and $7.5^{+0.6}_{-0.7} \cdot 10^{-13} \textrm{ergs}\ \textrm{cm}^{-2}\ \textrm{s}^{-1}$, respectively. The $2$-$10$ keV luminosity is $1.8^{+0.2}_{-0.1} \cdot 10^{46} \textrm{ergs}\ \textrm{s}^{-1}$. In the past, PKS 1502+106 was observed to have a photon index $\Gamma$ varying between $1.4$ and $1.9$ and a $2$-$10$ keV flux in the range $4.9-6.54\ \cdot 10^{-13} \textrm{ergs}\ \textrm{cm}^{-2}\ \textrm{s}^{-1}$ \citep{George94, Akiyama03, Watanabe04}. An outburst was observed by Fermi and Swift \citep{Abdo10}, with $F_{0.3-10 keV} = 2.18\ \cdot 10^{-12} \textrm{ergs}\ \textrm{cm}^{-2}\ \textrm{s}^{-1}$, and $\Gamma = 1.54$. Our best-fit model parameters suggest that \textit{Suzaku} detected PKS 1502+106 during quite a high state, but not during an outburst (its light curve does not show any significant variability). However, the power-law index is consistent with a low state of the source. 
  This source falls within the \textit{PIN} field of view, thus it has a contribution to the \textit{PIN} data. The extrapolated flux in the 12-60 keV band is $2.5 \cdot 10^{-12} \textrm{ergs}\ \textrm{cm}^{-2}\ \textrm{s}^{-1}$. This component was included in the \textit{PIN} background.\\
  During the second observation, PKS 1502+106 falls off the CCD image of Mrk~841, but it remains in the \textit{PIN} field of view: we subtracted from \textit{PIN} data the same spectrum evaluated for the first observation.\\
  No bright X-ray sources are observed in the \textit{XIS} field of view of Mrk~509.\\

   \section {Data analysis}
   \subsection{Light curves}

   \begin{figure*}[pt!]
	 \centering
   \resizebox{\hsize}{!}{\includegraphics{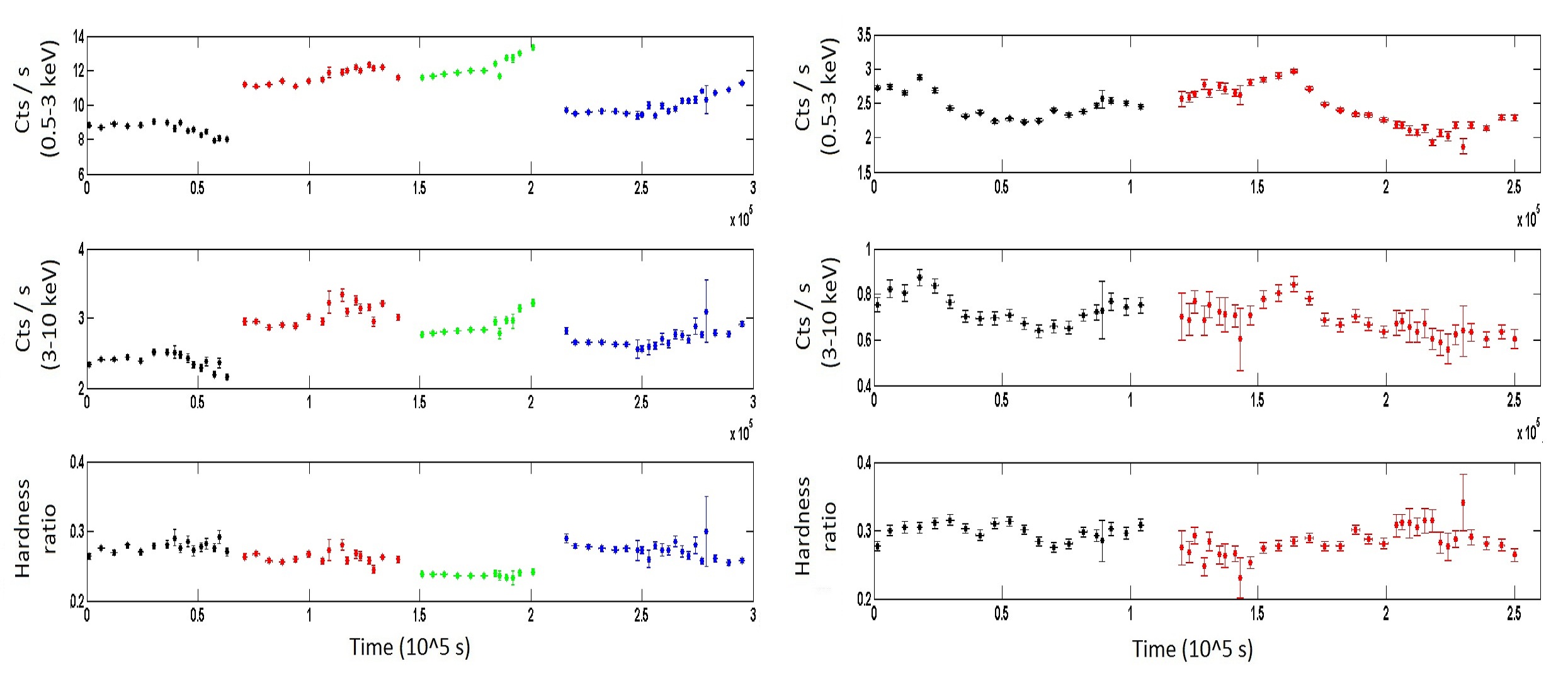}}
     \caption{Mrk~509 (left) and Mrk~841 (right) light curves. Different observations are shown in different colours (one-black, two-red, three-green, and four-blue). From top to bottom: $0.5$-$3$ keV light curve (in units of counts per second); $3$-$10$ keV light curve and hardness ratio ($3$-$10$/$0.5$-$3$). Gap between observations is arbitrary. }
     \label{Mrk509_lc}
	 \centering
   \resizebox{\hsize}{!}{\includegraphics{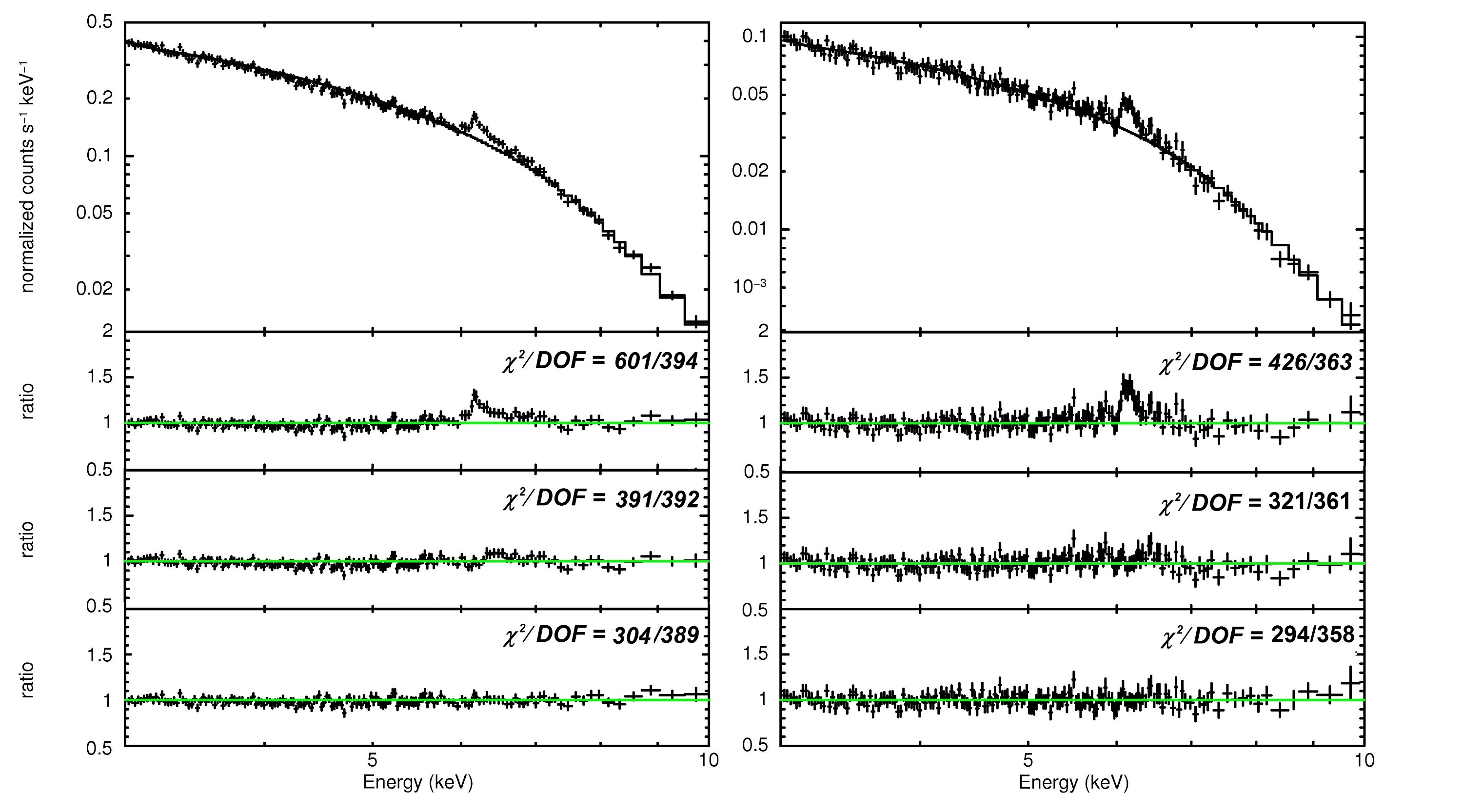}}
     \caption{\textit{XIS} data between $3$ and $10$ keV of the sum of Mrk~509 (left) and Mrk~841 (right) observations : data are first fitted with a power law (ignoring the $6$-$7$ keV energy band, upper panel and first ratio plot); then two narrow Gaussian lines (for the Fe $K\alpha$ and $K\beta$ emission lines, see section 3.2) are added to the fit (second ratio plot); a broad ($\sigma > 0$) Fe emission line is finally considered (third and last ratio plot).  Only \textit{FI} data are shown, as they have the highest resolution around the Fe $K\alpha$ emission line, but the fit was performed using both \textit{FI} and \textit{XIS}1 data. The energy on the abscissa is in the observer frame. The data were rebinned for clarity purposes.}
     \label{ratioplot}
   \end{figure*}
   
   \begin{table*}[ht!]
    \begin{center}   
   \caption{Best-fit parameter values for the model \textit{powerlaw + zgauss + zgauss + zgauss}, considering \textit{XIS} data between $3$ and $10$ keV. The three Gaussian functions model the Fe $K\alpha$ and $K\beta$ narrow emission lines (with $\sigma=0$) and the broad Fe K emission line. The Fe $K\beta$ emission line was fixed in energy ($E_{K\beta}=7.06$ keV) and in normalization (imposed equal to $0.15$ times the $K\alpha$ line normalization). $\Gamma$ is the power-law photon index; the power-law normalization $C_{\Gamma}$ is the number of photons $\textrm{keV}^{-1}\ \textrm{cm}^{-2}\ \textrm{s}^{-1}$ at $1$ keV; the line normalization parameters $C_{K\alpha}$ and $C_{br}$ are the integrated number of photons $\textrm{cm}^{-2}\ \textrm{s}^{-1}$ in the line; $\sigma_{br}$ is the line width expressed in keV; EqW is the line equivalent width expressed in eV; the flux $F$, evaluated between $3$ and $10$ keV, is given in units of $\textrm{ergs}\ \textrm{cm}^{-2}\ \textrm{s}^{-1}$. For the summed spectra, the results of a \textit{XIS} (above $3$ keV) + \textit{PIN} fit are reported as well (last line): the $R_{pexrav}$ parameter represents the \textit{pexrav} reflection scaling factor (see Section 3.3).}              % title of Table
   \label{po}      % is used to refer this table in the text
   \begin{tabular}{c c c c c c c c c c c c c}
   \hline
   \multicolumn{13}{c}{Mrk~509}\\
   \hline
   Obs. & $\Gamma$ & $C_{\Gamma}$ & $E_{K\alpha}$ & $C_{K\alpha}$ & $EqW_{K\alpha}$ & $E_{br}$ & $C_{br}$ & $\sigma_{br}$ & $EqW_{br}$ & $R_{pexrav}$ & $F$ & $\chi^2/DOF$\\[1pt]
   & & $(10^{-3})$ & $(keV)$ & $(10^{-5})$ & $(eV)$ & $(keV)$ & $(10^{-5})$ & $(keV)$ & $(eV)$ &  &$(10^{-11})$ & \\
   \hline
   1 & $1.80^{+0.05}_{-0.04}$ & $12.5^{+0.8}_{-0.6}$ & $6.42^{+0.09}_{-0.09}$ & $0.9^{+0.9}_{-0.8}$ & $16^{+18}_{-15}$ & $6.44^{+0.17}_{-0.16}$ & $6.2^{+3.6}_{-2.7}$ & $0.44^{+0.28}_{-0.17}$ & $126^{+79}_{-78}$ & & $3.45^{+0.02}_{-0.03}$ & $329/386$\\[1pt]
   2 & $1.79^{+0.04}_{-0.03}$ & $13.4^{+0.2}_{-0.3}$ & $6.47^{+0.06}_{-0.06}$ & $1.6^{+0.8}_{-0.7}$ & $28^{+19}_{-18}$ & $6.51^{+0.17}_{-0.18}$ & $4.6^{+3.0}_{-2.2}$ & $0.41^{+0.30}_{-0.18}$ & $\ 86^{+50}_{-61}$ & & $3.78^{+0.02}_{-0.02}$ & $375/366$\\[1pt]
   3 & $1.86^{+0.05}_{-0.04}$ & $14.4^{+0.4}_{-0.7}$ & $6.41^{+0.05}_{-0.05}$ & $1.7^{+0.9}_{-0.8}$ & $30^{+19}_{-17}$ & $6.55^{+0.25}_{-0.14}$ & $5.7^{+1.8}_{-1.8}$ & $0.44^{+0.29}_{-0.14}$ & $112^{+43}_{-45}$ & & $3.60^{+0.03}_{-0.02}$ & $344/352$\\[1pt]
   4 & $1.80^{+0.04}_{-0.03}$ & $12.1^{+0.7}_{-0.6}$ & $6.42^{+0.03}_{-0.03}$ & $1.8^{+0.7}_{-0.7}$ & $35^{+15}_{-15}$ & $6.71^{+0.21}_{-0.14}$ & $6.7^{+3.2}_{-2.4}$ & $0.52^{+0.25}_{-0.20}$ & $145^{+75}_{-65}$ & & $3.39^{+0.02}_{-0.02}$ & $396/378$\\[1pt]
   \hline
   2+3+4 & $1.84^{+0.03}_{-0.02}$ & $13.5^{+0.5}_{-0.4}$ & $6.43^{+0.03}_{-0.03}$ & $1.7^{+0.4}_{-0.5}$ & $31^{+6}_{-6}$ & $6.63^{+0.11}_{-0.09}$ & $6.6^{+2.2}_{-1.7}$ & $0.50^{+0.17}_{-0.12}$ & $135^{+48}_{-45}$ & & $3.50^{+0.01}_{-0.02}$ & $304/389$\\[1pt]
   \ \ \ \ +\textit{PIN} & $1.88^{+0.05}_{-0.05}$ & $14.0^{+0.6}_{-0.8}$ & $6.43^{+0.03}_{-0.03}$ & $1.7^{+0.5}_{-0.5}$ & $33^{+9}_{-14}$ & $6.69^{+0.16}_{-0.13}$ & $6.1^{+3.8}_{-1.4}$ & $0.56^{+0.34}_{-0.17}$ & $126^{+29}_{-67}$ & $0.4^{+0.2}_{-0.2}$ & $3.50^{+0.02}_{-0.02}$ & $319/405$\\[1pt]
   \hline
   \hline
   \multicolumn{13}{c}{Mrk~841}\\
   \hline
    Obs. & $\Gamma$ & $C_{\Gamma}$ & $E_{K\alpha}$ & $C_{K\alpha}$ & $EqW_{K\alpha}$ & $E_{br}$ & $C_{br}$ & $\sigma_{br}$ & $EqW_{br}$ & $R_{pexrav}$ &$F$ & $\chi^2/DOF$\\[1pt]
   & & $(10^{-3})$ & $(keV)$ & $(10^{-5})$ & $(eV)$ & $(keV)$ & $(10^{-5})$ & $(keV)$ & $(eV)$ &  &$(10^{-11})$ & \\
   \hline
   1 & $1.77^{+0.08}_{-0.05}$ & $3.77^{+0.20}_{-0.19}$ & $6.39^{+0.02}_{-0.03}$ & $1.0^{+0.3}_{-0.3}$ & $58^{+26}_{-22}$ &  $6.03^{+0.36}_{-0.31}$ & $2.7^{+2.7}_{-0.9}$ & $0.74^{+0.32}_{-0.31}$ & $156^{+105}_{-119}$ & & $1.10^{+0.03}_{-0.03}$ & $303/341$\\[1pt]
   2 & $1.75^{+0.06}_{-0.05}$ & $3.66^{+0.10}_{-0.25}$ & $6.38^{+0.04}_{-0.04}$ & $0.7^{+0.3}_{-0.4}$ & $39^{+32}_{-26}$ &  $6.26^{+0.18}_{-0.17}$ & $2.3^{+1.1}_{-0.9}$ & $0.40^{+0.22}_{-0.16}$ & $131^{+87}_{-80}$ & & $1.11^{+0.03}_{-0.03}$ & $259/323$\\[1pt]
   \hline
   1+2 & $1.77^{+0.05}_{-0.04}$ & $3.80^{+0.08}_{-0.21}$ & $6.38^{+0.02}_{-0.02}$ & $0.9^{+0.2}_{-0.2}$ & $50^{+19}_{-17}$ & $6.22^{+0.16}_{-0.19}$ & $2.7^{+0.9}_{-0.7}$ & $0.54^{+0.27}_{-0.15}$ & $150^{+66}_{-64}$ & & $1.11^{+0.03}_{-0.03}$ & $294/358$\\[1pt]
   \ \ \ \ +\textit{PIN} & $1.88^{+0.10}_{-0.05}$ & $4.18^{+0.60}_{-0.16}$ & $6.37^{+0.03}_{-0.01}$ & $0.9^{+0.2}_{-0.2}$ & $44^{+17}_{-18}$ & $6.14^{+0.24}_{-0.31}$ & $1.6^{+1.6}_{-0.9}$ & $0.54^{+0.35}_{-0.23}$ & $<158$ & $1.3^{+0.8}_{-0.3}$ & $1.11^{+0.03}_{-0.03}$ & $295/374$\\[1pt]
   \hline
   \end{tabular}
   \end{center} 
   \end{table*}

        \begin{figure*}[ht!]
	 \centering
   \resizebox{\hsize}{!}{\includegraphics[height=7.5cm]{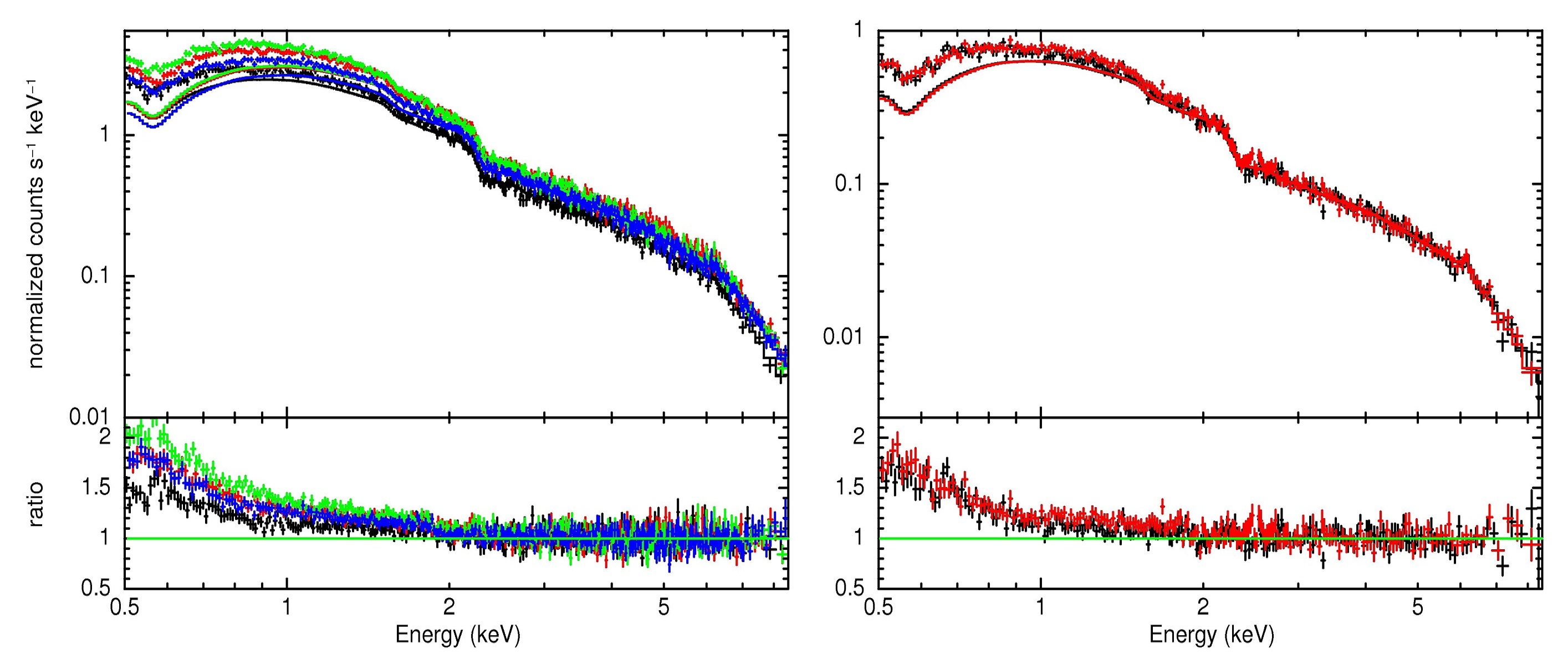}}
     \caption{Mrk~509 (left) and Mrk~841 (right) soft excess: data from $3$ to $10$ keV are fitted with a power law plus three Gaussian emission lines (see Table \ref{po}), and the model is then extrapolated to lower energies. Galactic absorption is taken into account. Only \textit{XIS1} data are shown, as they extend to $0.5$ keV, but the analysis was performed using both \textit{FI} and \textit{XIS}1 data; the colour code for the different observations is as in Fig. \ref{Mrk509_lc}. The energy on the abscissa is in the observer frame. The data were rebinned for clarity purposes.}
     \label{soft1}
   \end{figure*}

    \textit{XIS} light curves of Mrk~509 and Mrk~841 are shown in Fig. \ref{Mrk509_lc}. Owing to the orbit of \textit{Suzaku}, the targets, periodically occulted by Earth, cannot be observed continuously and light curves are composed by several blocks of a few kiloseconds each. As no significant variability appears on this timescale, we decided to rebin the data to one bin per observation window. Light curves were extracted for two energy bands, a soft ($0.5$-$3$ keV) one and a hard ($3$-$10$ keV) one, and a count-rate hardness ratio ($h = hard/soft$) was evaluated. The \textit{PIN} light curves are always compatible with a constant flux.\\
    
    During the first observation of Mrk~509, the source was in the \textit{HXD} aim point, which provides a smaller ($\approx 10\%$) effective area and a lower number of counts per second than for the \textit{XIS} aim point (which was used for the three other observations). For this reason, the lightcurve of observation (obs.) 1 had to be corrected for this systematic error when compared to the other observations. Taking into account this effect, Mrk~509 displays a variability of $\approx25\%$ between the four observations in the soft energy band, but within a single observation the flux variation is only $11$-$14\%$, reaching $20\%$ only during the last observation. In the hard energy domain, the variability is $\approx16\%$ between observations and $\approx15\%$ (but $20\%$ during the fourth observation) on a $10^5$s timescale.
    
  The value of the hardness ratio during obs. 1, 2, and 4 is compatible with a constant (the mean values are $h=0.27^{+0.01}_{-0.01}$, $0.26^{+0.01}_{-0.01}$ and $0.27^{+0.01}_{-0.01}$, respectively). During the third observation $h$ has instead a lower value ($0.24^{+0.01}_{-0.01}$). \citet{Miyazawa09} analysed the temporal and spectral variability of a sample of 36 AGN observed by \textit{Suzaku}: the four observations of Mrk~509 we present here, were studied by these authors, and a correlation plot between the $0.5$-$2$ keV and the $2$-$10$ keV light curves is shown (see Fig.3 in their paper). They found a correlation between the fluxes in the soft and the hard energy band, consistent with the nearly constancy of the hardness ratio, though the lower value of $h$ during the third observation increases the dispersion in the correlation plot.\\
      
    The two Mrk~841 observations have an almost equal mean flux value, but the source shows variability on timescales of $10^5$s ($26\%$ and $46\%$ in the soft X-ray domain, and $32\%$ and $42\%$ in the hard X-ray domain, for the two observations respectively). For this source, too, the hardness ratio is nearly constant, and we measure a mean value of $h=0.29^{+0.02}_{-0.02}$.\\
    
    Given the low variability observed in Mrk~509 and Mrk~841 light curves, and to improve the signal-to-noise ratio, we study in the following the summed spectra as well as individual observations for phenomenological models.\\

   \subsection{Soft excess}

   We start the analysis considering the \textit{XIS} $3$-$10$ keV energy band. We model the data using a power law plus two narrow Gaussian functions (model \textit{zgauss} in \textit{XSpec}, assuming $\sigma=0$) for the Fe K$\alpha$ and K$\beta$ narrow emission lines (assuming $E_{K\beta}=7.06$ keV and imposing its normalization factor $C_{K\beta}$ equal to $0.15\ C_{K\alpha}$ \citep{Palmeri03}). An excess of emission is clearly observed around $6.4$ keV (see Fig. \ref{ratioplot}) and adding a broad emission line significantly improves the fit (F-test probability equal to $4\cdot10^{-21}$ and $7\cdot10^{-7}$ for the summed Mrk~509 and Mrk~841 data set, respectively). Table \ref{po} shows the best-fit results. The detection of a Fe broad emission line in \textit{Suzaku} observations is consistent with the results showed in the FERO (Finding Extreme Relativistic Objects) survey \citep{delaCalle10}, which studied, using \textit{XMM-Newton} data, the relativistic Fe K$\alpha$ emission lines in Seyfert 1 galaxies, including Mrk~509.\\ \citet{Ponti09} studied in detail the Fe K complex in Mrk~509, using the same \textit{Suzaku} observations presented here: in particular, they showed that a broad Fe line (with parameters consistent with our values given in Table 2) represents the best-fitting model, compared to a possible blend of narrow Fe lines.\\For Mrk~841, we performed the same analysis as \citet{Ponti09}, testing the possibility that a blending of H-like and He-like lines of Fe might be responsible for the broad line observed : however, this model (two narrow Gaussian lines, with energies fixed to $6.67$ and $6.97$ keV, plus the $6.4$ and $7.06$ Fe lines, as described above) does not significantly improve the goodness of the fit (F-test probability equal to $0.1$ for the summed data-set).\\ For Mrk~509, we observe variability in the $3$-$10$ keV flux (comprised between $3.39$ and $3.78\cdot 10^{-11}\ \textrm{ergs}\ \textrm{cm}^{-2}\ \textrm{s}^{-1}$), while the slope is consistent, within the errors, with a constant (the two extreme values are $1.79^{+0.04}_{-0.03}$ and $1.86^{+0.05}_{-0.04}$ for the second and the third observation, respectively). The best-fit results of the two Mrk~841 observations are quite similar ($\Gamma = 1.75$-$1.77$ and $F=1.10$-$1.11\cdot 10^{-11}\ \textrm{ergs}\ \textrm{cm}^{-2}\ \textrm{s}^{-1}$), as discussed in the previous section.
   
   It is useful to compare these results to past observations. Using \textit{BeppoSAX}, \citet{DeRosa04} detected Mrk~509 in two quite different states, at $F_{2-10\ keV}=2.7$ and $5.7~\cdot~10^{-11}\ \textrm{ergs}\ \textrm{cm}^{-2}\ \textrm{s}^{-1}$ with $\Gamma=1.80$-$1.59$ (for comparison the $2$-$10$ keV flux of Mrk~509 during the four \textit{Suzaku} observations varies between $4.3$ and $4.8\cdot 10^{-11}\ \textrm{ergs}\ \textrm{cm}^{-2}\ \textrm{s}^{-1}$, setting these observations in an intermediate state). \textit{XMM-Newton} \citep{Smith07} observed the source in a even lower state at $F_{0.5-10\ keV}=2.6 \cdot 10^{-11}\ \textrm{ergs}\ \textrm{cm}^{-2}\ \textrm{s}^{-1}$. 
   
   Mrk~841 was observed by \textit{XMM-Newton} (see P07) at $F_{3-10\ keV} = 0.8$-$1.3 \cdot 10^{-11}\textrm{ergs}\ \textrm{cm}^{-2}\ \textrm{s}^{-1}$ with $\Gamma=1.30$-$1.95$. Compared to these values, the source was in an intermediate state during \textit{Suzaku} observations; in the past, Mrk~841 was seen in a quite low state by \citet{Pounds94} (simultaneous \textit{Ginga} and \textit{ROSAT} observations) with $F_{2-10\ keV} = 7.9 \cdot 10^{-12}\ \textrm{ergs}\ \textrm{cm}^{-2}\ \textrm{s}^{-1}$.\\  
   
   Extrapolating the power-law fit between $3$ and $10$ keV  to low energies, an excess of photons is observed: in Fig. \ref{soft1}, we report \textit{XIS}1 spectra for Mrk~509 and Mrk~841, and for each observation; the bottom panel shows the data-to-model ratio. The Galactic absorption has been taken into account following \citet{Dickey90}, using $N_H = 4.11 \cdot 10^{20} \textrm{cm}^{-2}$ for Mrk~509 and $N_H = 2.34 \cdot 10^{20} \textrm{cm}^{-2}$ for Mrk~841. The soft excess shape is remarkably similar for the two Seyfert galaxies, and for Mrk~509 spectral variability is observed at low energies.\\ 
   As a first qualitative study, we model the $0.5$-$10$ keV spectrum with a broken power law (plus three Gaussian emission lines, with parameters fixed as in Table \ref{po}, and Galactic absorption, fixed at values given above) to evaluate the soft excess slope ($\Gamma_s$) and energy break $E_b$. As shown in Table \ref{bkpo}, this model does not provide a good fit to our data : the soft spectrum between $0.5$ and $3$ keV clearly displays some curvature and cannot be described by a power law.\\
   \subsection{Hard excess}
   Extrapolating the $3$-$10$ keV best-fit power law to high energies, using \textit{PIN} data, a hard excess can also be seen. This feature is well-explained by introducing a distant reflection by cold material, which accounts at the same time for the Fe $K\alpha$ narrow emission line and the high energy bump. To quantify the amount of reflected emission required, we fitted the summed \textit{PIN} and \textit{XIS} (above $3$ keV) data with a power law (plus three Gaussian lines, as described in the previous section) plus \textit{pexrav} \citep{pexrav}, a model that describes reflection from neutral material. The \textit{pexrav} free parameters are the reflection scaling factor $R_{pexrav}$ (which is equal to 1 for a source isotropically illuminating the disc); the illuminating power-law high-energy cut-off $E_c$ (fixed at $100$ keV); the abundances (fixed to equal solar values) and the disc inclination angle $\theta$ (fixed at $30$ degrees). For the summed spectra, we obtain $R_{pexrav} = 0.4^{+0.2}_{-0.2}$ and $ 1.3^{+0.8}_{-0.3}$ for Mrk 509 and Mrk 841, respectively. Adding the neutral reflection component slightly modifies the best-fit parameters of the broad Fe emission line, in particular its equivalent width. Best-fit parameters are given in Table 2. \textit{Pexrav} has the advantage that gives directly the information about the reflected fraction $R_{pexrav}$, but does not evaluate the narrow Fe $K\alpha$ emission line produced. We then decided to use \textit{reflext}, a modified (and not public) version of \textit{reflion} model (an older version of \textit{reflionx}  \citep{Ross99, RossFabian}; see section $3.6$). Unlike \textit{reflion}, in which the ionisation parameter $\xi$ is comprised between $10$ and $10^4\ \textrm{erg cm s}^{-1}$, in \textit{reflext} the $\xi$ parameter space was extended as low as $1\ \textrm{erg cm s}^{-1}$. Using \textit{reflext}, the evaluation of the reflected fraction is less accessible, but the Fe $K\alpha$ narrow emission line and the high energy bump are intrinsically tied.\\
     In the following, we add \textit{reflext} with parameters $Fe/Fe_{\odot}=1$, $\Gamma_{reflext}=\Gamma$, and $\xi = 1\ \textrm{erg cm s}^{-1}$. Its normalization parameter $C_{reflext}$ was fixed to ensure that the flux of the model in the narrow Fe $K\alpha$ emission line reproduced the observed line flux. As the intensity of the reflected emission depends on the slope of the illuminating continuum, we evaluated the value of $C_{reflext}$ that reproduces the observed line flux as a function of $\Gamma$; we fitted the result with a polynomial function, and we fixed the \textit{reflext} normalization as a function of $\Gamma$.\\
      
   \subsection{Fe $K\alpha$ narrow emission line in Mrk~841}
   
\begin{figure}[pt]
	 \centering
   \resizebox{\hsize}{!}{\includegraphics[height=5cm]{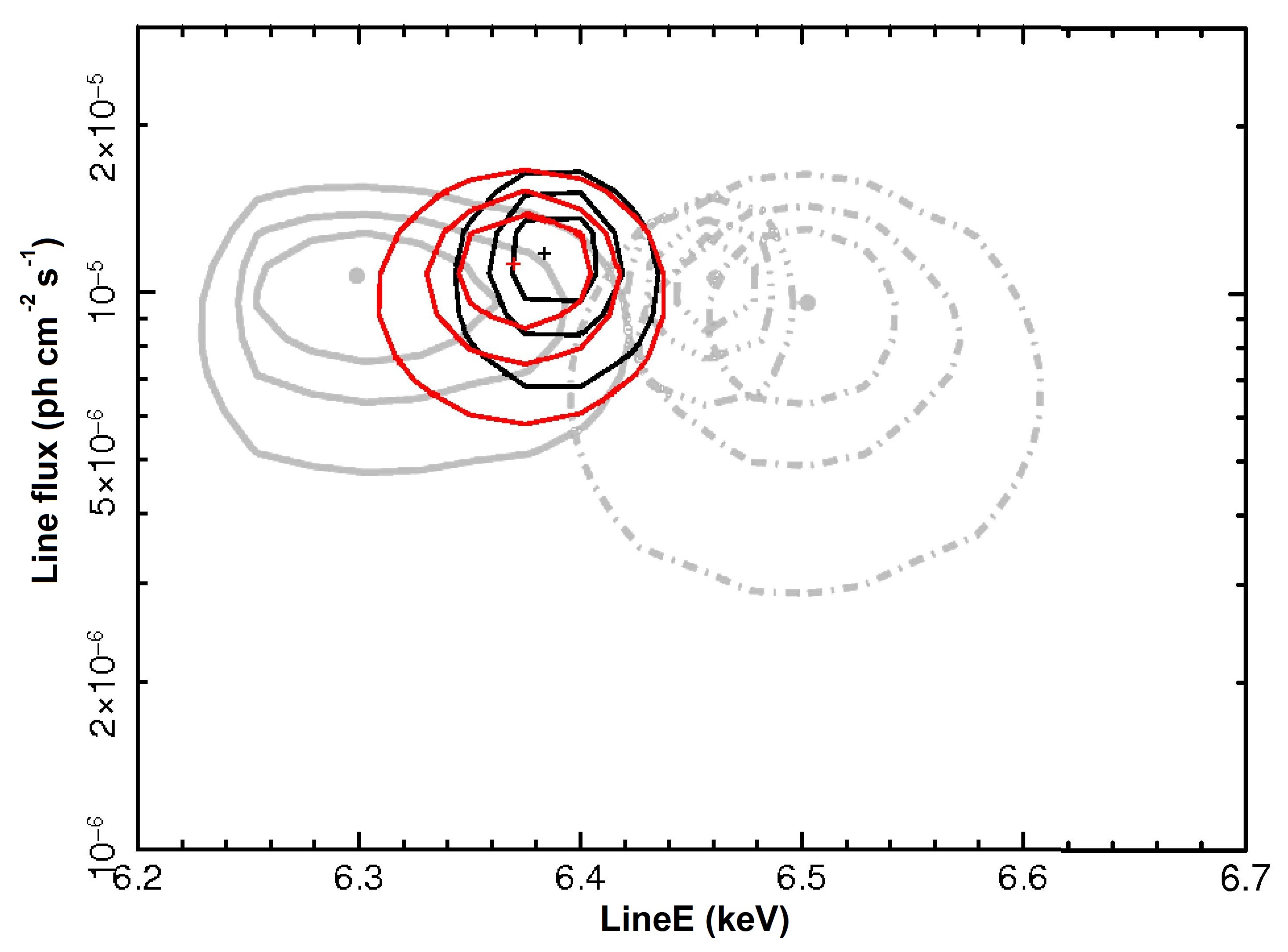}}
     \caption{Contour plots (in black for obs.1, and in red for obs.2) of line flux versus line energy (source frame) for the narrow emission Gaussian line in Mrk~841. The three confidence levels are for $68$, $90$, and $99\%$, respectively. In grey, we superimpose the contour plots of 2001-2005 \textit{XMM/Newton} observations (see P07, Fig. 9). }
     \label{narrow}
   \end{figure}

   We present here the analysis of the Fe $K\alpha$ emission line in Mrk~841. \citet{Ponti09} already presented a thorough study of the iron K complex in Mrk~509, using the same \textit{Suzaku} observations we analysed.\\
    A narrow ($\sigma=0$) Fe $K\alpha$ emission line was observed in previous Mrk~841 observations. In particular, P07 claimed that this narrow component is roughly constant in time, in agreement with a remote reflection scenario. We tested this hypothesis fitting the $3$-$10$ keV \textit{XIS} data with a power law plus a \textit{narrow} Gaussian emission line, with $\sigma$ fixed to zero, as done by P07. In Fig. \ref{narrow}, we report the contour plots of line flux versus line energy, for the two \textit{Suzaku} observations, and we compare them to the results of P07 (\textit{XMM/Newton}, observations from 2001 to 2005). The line flux agrees with the findings of P07, and no variability is observed. Line energy, in the source frame, is consistent with iron $K\alpha$ for both observations.\\
  
       \begin{table}
   \caption{Best-fit parameter values for the broken power-law model. \textit{XIS/FI} data between $0.6$ and $10$ keV and \textit{XIS1} data between $0.5$ and $8.5$ keV were considered. The power-law normalization $C_h$ is the number of photons $\textrm{keV}^{-1}\ \textrm{cm}^{-2}\ \textrm{s}^{-1}$ at $1$ keV.}              % title of Table
   \label{bkpo}      % is used to refer this table in the text
   \begin{tabular}{c c c c c c}
   \hline
   \multicolumn{6}{c}{Mrk~509}\\
   \hline
   Obs. & $\Gamma_s$ & $E_b$ (keV)& $\Gamma_h$ & $C_h(10^{-3})$ & $\chi^2/DOF$ \\[1pt]
   \hline
   1 & $2.24^{+0.08}_{-0.05}$ & $1.02^{+0.07}_{-0.09}$ & $1.87^{+0.01}_{-0.01}$ &  $13.8^{+0.2}_{-0.4}$ & $743/726$ \\[1pt]
   2 & $2.16^{+0.02}_{-0.02}$ & $1.83^{+0.09}_{-0.08}$ & $1.86^{+0.01}_{-0.01}$ &  $18.0^{+0.1}_{-0.1}$ & $849/677$ \\[1pt]
   3 & $2.29^{+0.01}_{-0.01}$ & $1.74^{+0.06}_{-0.07}$ & $1.93^{+0.01}_{-0.01}$ &  $19.3^{+0.1}_{-0.1}$ & $902/652$ \\[1pt]
   4 & $2.12^{+0.01}_{-0.01}$ & $1.96^{+0.09}_{-0.10}$ & $1.84^{+0.01}_{-0.01}$ &  $15.6^{+0.1}_{-0.1}$ & $936/689$ \\[1pt]
   2+3+4 & $2.19^{+0.01}_{-0.01}$ & $1.77^{+0.04}_{-0.04}$ & $1.89^{+0.01}_{-0.01}$ &  $17.3^{+0.1}_{-0.1}$ & $1248/721$ \\[1pt]
   \hline
   \hline
   \multicolumn{6}{c}{Mrk~841}\\
   \hline
   Obs. & $\Gamma_s$ & $E_b$ (keV) & $\Gamma_h$ & $C_h(10^{-3})$ & $\chi^2/DOF$ \\[1pt]
    \hline
   1 & $2.44^{+0.06}_{-0.08}$ & $1.02^{+0.06}_{-0.04}$ & $1.82^{+0.01}_{-0.01}$ &  $4.09^{+0.07}_{-0.07}$ & $686/651$ \\[1pt] 
   2 & $2.40^{+0.09}_{-0.08}$ & $0.96^{+0.06}_{-0.04}$ & $1.87^{+0.01}_{-0.01}$ &  $4.21^{+0.12}_{-0.13}$ & $604/613$ \\[1pt]
   1+2 & $2.44^{+0.06}_{-0.06}$ & $1.00^{+0.04}_{-0.03}$ & $1.85^{+0.01}_{-0.01}$ &  $4.18^{+0.08}_{-0.08}$ & $804/694$ \\[1pt]
   \hline
   \end{tabular} 
   \end{table} 
   
    In the following, we concentrate on the broad-band spectral fitting of Mrk~509 and Mrk~841 data with three different physical models : a double Comptonisation, a blurred reflection, and a smeared absorption. As only weak flux variability is observed between individual Mrk~509 observations, and as no significant variability is observed between individual Mrk~841 observations, we perform and discuss the broad-band spectral analysis for the summed spectra only (unless explicitly mentioned).
   
   \subsection{Double Comptonisation model}

     In the standard disc-corona scenario \citep{Haardt93}, the power-law high-energy emission is explained as a Comptonisation of thermal photons emitted by the black hole accretion disc. Electrons producing this inverse Compton effect have a temperature $kT_e \approx 100$ keV and a low optical depth, $\tau \approx 1$. One possible way of easily explaining the soft excess is that there is a second warm Comptonisation region, of lower temperature and higher optical depth, that modifies the black-body emission from the accretion disc, scattering UV photons up to soft X-ray energies. A physical interpretation of this ''second corona'' might be a warm skin on the disc surface \citep{Janiuk01}, or a transition region between an outer disc and an optically thin inner flow \citep{Magdziarz98}. We used \textit{NTHComp} \citep{Zycki99}, which is one of the \textit{XSpec} models, to simulate Comptonised spectra: this model solves the Kompaneets equation, and the output spectrum is parametrised by the electron temperature $kT_e$ and the spectral index $\Gamma$; seed photons come from an accretion disc spectrum, parametrised by $kT_{disc}$, the temperature at the inner radius. The electron temperature of the corona producing the high energy emission (subscript \textit{h}), $kT_h$, can be constrained only via the high energy cut-off: since we do not observe this cut-off in our data, we fixed it at $100$ keV. For the soft Comptonisation region (subscript \textit{s}), we cannot fit at the same time $\Gamma_s$ and $kT_s$, and, following GD04, we set $\Gamma_s=2$.\\
The exact value of $kT_{disc}$ does not affect the fit as long as it is lower than our energy domain, thus we chose $kT_{disc}=45$ eV. We added \textit{reflext} to our model to describe the reflection from distant material, with a normalization parameter fixed as described above, plus two Gaussian functions for the Fe $K\beta$ and the Fe $K_{broad}$ emission lines, as in Table \ref{po} (for the \textit{XIS} + \textit{PIN} fit). Our best-fit model results are given in Table \ref{nthcomp_zxipcf}.\\ 
       
     The double Comptonisation model was studied in detail by GD04, who fitted a sample of 26 AGN showing that the cold Comptonisation region has a remarkably constant temperature <$kT_s$>$=0.12$ keV with $\sigma = 0.02$ keV, despite large variations in black hole mass and AGN luminosity. They also used the model \textit{NTHComp}. The values we found for Mrk~509 and Mrk~841 during Suzaku observations agree with their statistics. We recall that the exact value of $kT_s$ depends on the assumption made for $\Gamma_s$: the only physical parameter that can be constrained (when the high energy cut-off is not observed) is the Comptonisation parameter $y$, defined as $4kT/(mc^2)\cdot Max(\tau, \tau^2)$, where $\tau$ is the plasma optical depth. The $\Gamma$ parameter represents the asymptotic power-law function that describes the Comptonised spectrum, and can be expressed as $(2.25 + 4/y)^{0.5} - 0.5$, which is equal to $2$ for $y=1$. The most important result obtained by GD04 is not the exact value of $kT_s$, but its constancy in AGN samples.\\
     Following GD04, we evaluated the soft-excess strength, $R_{GD}$, defined as the ratio of the unabsorbed $0.3$-$2$ keV flux in the cool to that in the hot Comptonised components: it is equal to $0.23^{+0.04}_{-0.05}$ for Mrk~509 and $0.28^{+0.08}_{-0.04}$ for Mrk~841, placing our sources in the lowest part of the GD04 sample in terms of soft-excess strength.\\
      
     \subsubsection{Additional contribution of the warm absorber}  
     
   In terms of the $\chi^2$ values, the double Comptonisation model does not provide a good fit to the data. For both sources, the presence of a warm absorber was confirmed, and its features could explain the residuals of the fit. The most complete studies of the warm absorber of these two objects were done using \textit{XMM/RGS} data by \citet{Smith07} and \citet{Detmers10} for Mrk~509, and by \citet{Longinotti10} for Mrk~841.\\ Smith et al. detected a three-phase warm absorber in Mrk~509, consisting of a low ionisation component (log~$\xi$ $\approx$ $0.89$  and column density $N_H\approx 7.9 \cdot 10^{20} \textrm{cm}^{-2}$), an intermediate ionisation phase (log~$\xi \approx 2.14$, $N_H\approx 7.5 \cdot 10^{20} \textrm{cm}^{-2}$), and a high ionisation one (log~$\xi \approx 3.26$,  $N_H\approx 55 \cdot 10^{20} \textrm{cm}^{-2}$). Detmers et al. confirmed the presence of three different components in the warm absorber: they found a low ionisation phase (log~$\xi \approx 0.6$, $N_H\approx 1.0 \cdot 10^{20} \textrm{cm}^{-2}$), an intermediate ionisation phase (log~$\xi \approx 1.95$, $N_H\approx 10.5 \cdot 10^{20} \textrm{cm}^{-2}$), and a high ionisation phase (log~$\xi \approx 3.2$, $N_H\approx 80.0 \cdot 10^{20} \textrm{cm}^{-2}$).\\ A multi-phase warm absorber was also found in Mrk~841, by Longinotti et al., with an intermediate ionisation component (log~$\xi \approx 1.5$-$2.2$, $N_H\approx 12$-$39 \cdot 10^{20} \textrm{cm}^{-2}$) and a higher ionisation phase (log~$\xi=2.8$-$3.3$  and $N_H\approx 76$-$300 \cdot 10^{20} \textrm{cm}^{-2}$).\\ We tested the presence of the warm absorber using the model \textit{zxipcf} \citep{Reeves08}, which is another of the \textit{XSpec} models, which has as free parameters the ionisation parameter log~$\xi$, the column density $N_H$, the covering factor $f$, and the redshift $z$. We started our study by multiplying the double Comptonisation model by one \textit{zxipcf} component, with $f=1$ and $z=z_{object}$. For Mrk~509, we found that log~$\xi_{WA} = 2.19^{+0.05}_{-0.05}$ and $N_{H,WA} = 17^{+2}_{-2}\cdot 10^{20} \textrm{cm}^{-2}$, values that are consistent with the intermediate ionisation phase found by Smith et al. and Detmers et al.. For Mrk~841, we found that log~$\xi_{WA} = 2.3^{+0.1}_{-0.1}$ and $N_{H,WA} = 17^{+4}_{-4} \cdot 10^{20} \textrm{cm}^{-2}$, which is consistent with the intermediate ionisation phase in Longinotti et al.. Adding \textit{zxipcf} always improves the goodness of the fit, as is confirmed by a F-test: the null hypothesis probabilities were found to be $7\cdot10^{-44}$ (for Mrk~509) and $1\cdot10^{-13}$ (for Mrk~841).\\
   We then tried to add a second absorber (subscript $2$). For Mrk~509, we found that a two-phase warm absorber is statistically preferred (the associated F-test probability is $5\cdot 10^{-5}$) and we measured log~$\xi_{WA,2} = 2.7^{+0.4}_{-0.1}$, $N_{H,WA,2} = 7^{+9}_{-2} \cdot 10^{20} \textrm{cm}^{-2}$, log~$\xi_{WA,1} = 1.7^{+0.1}_{-0.1}$, and $N_{H,WA,1} = 8^{+2}_{-1} \cdot 10^{20} \textrm{cm}^{-2}$. 
   For the high ionisation phase, \textit{XMM/RGS} data also showed a warm-absorber flow velocity blueshifted relative to the galaxy: we then fixed $z_{WA,2}$ to the value found by Detmers et al. ($z_{WA,2} = 0.0334$, corresponding to $v=-290\ \textit{km s}^{-1}$).
   For Mrk~841, adding a second absorber does not improve the goodness of the fit and the absorber parameters are poorly constrained: we then fixed log~$\xi_{WA,2} = 3.3$, which is equal to the value measured by Longinotti et al. during the 2001 \textit{XMM} observation of Mrk~841.\\Our best-fit results are given in Table \ref{nthcomp_zxipcf} and plotted in Figs. \ref{Mrk509_ldata} and \ref{Mrk841_ldata} (left panel). We considered whether there is a three-phase warm absorber in Mrk~509, but the goodness of the fit worsens, even fixing the warm absorbers parameters at the values measured by \textit{XMM/RGS}.\\ 
   
The \textit{Zxipcf} model can describe partial covering absorbers, with $f$$<$$1$. We tested the effect of this parameter on a one-phase warm absorber. For Mrk~509, leaving $f$ free to vary, it decreases to $0.3$, while $N_{H,WA}$ increases, reaching $3 \cdot 10^{22} \textrm{cm}^{-2}$; the new fit is statistically better than the one with $f=1$ (the null hypothesis probability is equal to $1\cdot10^{-7}$). For Mrk~841, the best-fit result is always consistent with $f=1$ (we measure $f>0.7$). Despite the better fit obtained with $f$$<$$1$ for Mrk~509, we decided to use and study the $f=1$ case in order to compare our results to those obtained with the \textit{XMM/RGS} data (and for consistency between the analysis of our two sources). In any case, there is a degeneracy between $f$ and $N_H$, and leaving $f$ free affects the values of $\xi$ and $T_s$ and, consequently, the physical interpretation of our results. In the next sections, we always use $f=1$, with a reminder that, for Mrk~509, allowing $f$ free to vary always leads to a better fit.\\

In both sources, \textit{XMM/RGS} data revealed an OVII emission line at $E \approx 0.56$ keV.
We tested for the presence of this emission by adding a narrow ($\sigma=0$) Gaussian line to the model (the fit was performed using data below $10$ keV, without \textit{reflext}, which produces an OVII emission line as well (see below)).\\ For Mrk~509, we obtained a significantly better fit (F-test probability = $4\cdot10^{-9}$) for the fourth observation, while for the other observations and the summed spectra the goodness of the fit did not improve. However, at $0.56$ keV we do not have the \textit{XIS/FI} detectors (which start at $0.6$ keV), and we are at the energy edge of \textit{XIS1}: the energy resolution is quite low (the emission line is defined by one bin only) and we cannot confirm whether this statistically significant improvement to the fit corresponds to a true emission-line detection.\\
For Mrk~841, adding a Gaussian OVII emission line does not improve the goodness of the fit in either any individual observation or the summed spectrum.\\
The origin of this emission line is thought to be the warm gas itself but we do also expect OVII emission from colder reflection. In particular, \citet{Ebrero2010} showed, for the Seyfert 1 Mrk 279, that the \textit{reflext} code (for log~$\xi<1.43$) produces an OVII line emission flux that exceeds the observed OVII line flux by a factor of 10 or more. For Mrk~509 and Mrk~841, this excess does not exist: the line flux produced by \textit{reflext} (with log~$\xi=0$) is equal to $2.1 \cdot 10^{-4}\ \textrm{cm}^{-2}\ \textrm{s}^{-1}$ for Mrk~509 (compared to $2.7 \cdot 10^{-4}\ \textrm{cm}^{-2}\ \textrm{s}^{-1}$ found by Detmers et al.), and to $7.2 \cdot 10^{-5}\ \textrm{cm}^{-2}\ \textrm{s}^{-1}$ for Mrk~841 (compared to $(6.4 \pm 5.3) \cdot 10^{-5}\ \textrm{cm}^{-2}\ \textrm{s}^{-1}$ found by Longinotti et al. in the January 2005 \textit{XMM/RGS} observation).\\ 

   GD04 performed their fits without testing for the presence of a warm absorber in their data. Looking at Table \ref{nthcomp_zxipcf}, we can see that, after including \textit{zxipcf}, $kT_{s}$ increases from $0.13$-$0.14$ keV to $0.19$-$0.21$ keV, becoming only marginally consistent with GD04 results.\\

   \subsection{Blurred reflection model}
   
   The constancy and universality of the soft excess temperature is difficult to explain in a double Comptonisation scenario, for objects with different black hole masses (GD04). It is instead expected if the soft excess origin is linked to atomic processes. If the reflection from the accretion disc is ionised, and if the emission lines are broadened by a relativistic velocity near the black hole, this component might also explain the soft excess. We used the \textit{reflionx} model \citep{Ross99, RossFabian}, convolved with the \textit{kdblur} model \citep{Laor91}, to describe the disc ionised reflection. For the relativistic blurring, we assumed that $R_{in}=1.235\ R_G$ (the last stable orbit for a near extremal Kerr black hole with parameter $a=0.998$, \citep{Thorne74}), $R_{out}=100\ R_G$, $\theta = 30$ degrees, and disc emissivity index $\beta = 3$ \citep{Laor91}; for \textit{reflionx}, we fixed the iron abundance to equal the solar value, leaving $\xi$ and the normalization free. Following what we did in the previous section, we added a neutral reflection from distant material (\textit{reflext}, with normalization parameter fixed as a function of $\Gamma$, as described in Section 3.3), a narrow Gaussian line (for the Fe $K\beta$ emission line), and a two-phase warm absorber totally covering the source (\textit{zxipcf}), with an ionisation parameter and column density left free to vary within the error bars of the values found using a double comptonisation continuum (see Section 3.5.1). A broad emission feature around $6.4$ keV is expected from the blurred reflection on the accretion disc, so no broad emission line was included in the model. Best-fit model results are given in Table \ref{reflionx-beta-zxipcf}.\\
   The R parameter, which gives information about the reflected flux fraction, as defined by C06, is the ratio of the unabsorbed reflected component (\textit{reflionx}) flux to the total (excluding \textit{reflext}) unabsorbed flux in the $0.3$-$12$ keV energy band. For a flat disc illuminated by an isotropic source, we evaluated a value of $R\approx0.5$ for a high ionisation state (log~$\xi>3.5$), while for lower ionisations the reflection fraction decreases, reaching $0.4$ for log~$\xi=3$, $0.2$ for log~$\xi = 2$, and $0.1$ for log~$\xi=1.5$.\\ This model is widely used in the literature: C06 analysed a sample of 34 type 1 AGN, fitting their \textit{XMM-Newton} observations with the \textit{reflion} model (an older version of \textit{reflionx}). For the $R$ parameter, they found that one third of the sources have no power-law component ($R=1$), while in the rest of the sample $R$ is almost constantly distributed between $0.25$ and $0.8$. Our values, which range between $0.15$ and $0.18$, are quite low compared to this result. In particular, Mrk~841 is included in their sample, and they found that $R=0.67\pm0.17$. This value differs from what was found by P07, in a dedicated paper on \textit{XMM-Newton} observations, including the one studied by C06, for which they found that $R=0.3$ (in their paper, P07 evaluated that $R'=0.3$, where $R'$ is the reflected fraction evaluated in the energy band $0.1$-$1000$ keV, when including warm absorption; we reevaluated $R$ using their best-fit model parameters). In our model, we froze all \textit{kdblur} parameters. We tried to allow $\beta$ to vary, but its best-fit value is always consistent with $3$. C06 performed their fits leaving all, apart from $R_{out}$, of the \textit{kdblur} parameters free to vary: with such a large number of parameters, best-fit model values become less well-constrained, with several parameters pegged at either lower or higher limits (as can be seen in C06, Table 3) and a physical interpretation of these results can be misleading. The results of P07 for their \textit{XMM-Newton} observations of Mrk~841 were $1.59\leq \Gamma \leq 2.45$, $2\leq log\ \xi\leq 2.5$, and $\beta\approx4$ (but in one observation they found that $\beta>8.6$), with a reflection fraction $R'$ varying between $0.30$ and $0.43$.\\  

The best-fit model result for Mrk~509 is not good ($\tilde{\chi^2} = 1.25$) and some residuals are clearly observed between $6$ and $7$ keV (see Fig. \ref{reflionratio}). We first tried to vary the inner radius and the inclination angle of the accretion disc, significantly improving the fit (F-test probability = $1\cdot10^{-4}$) and measuring $R_{in}<1.7\ R_G$ and $\theta = 38^{+2}_{-3}$ degrees. However, this cannot take into account all the residuals observed near the Fe K$\alpha$ line. An excess of emission in the Fe K$\alpha$ complex might be caused by an iron abundance higher than the solar value: we tried to vary this parameter but the best-fit model results are always consistent with $\textrm{Fe}/\textrm{Fe}_{\odot} = 1$. Adding a broad Gaussian line, the fit is significantly improved (the null hypothesis probability is $10^{-29}$). We measured  $E_{0, broad} = 6.9^{+0.1}_{-0.1}$ keV, $C_{broad} = 10^{+1}_{-1}\cdot10^{-5}\ \textrm{photons cm}^{-2}\ \textrm{s}^{-1}$, and $\sigma_{broad} = 0.8^{+0.3}_{-0.2}$ keV, with an equivalent width of $255^{+85}_{-66}$ eV (see Table \ref{reflionx-beta-zxipcf} and Fig.\ref{Mrk509_ldata}, middle panel).\\ 
For Mrk 841, a good fit to the data was achieved using the first model (all \textit{kdblur} parameters were kept fixed at the values described above, and there was assumed to be no broad emission line), but the fit improved when $\theta$ was allowed free to vary, and we found that $\theta = 41^{+3}_{-4}$ degrees (with a F-test probability equal to $4\cdot10^{-4}$). Adding a broad emission line at the Fe K complex energy does not improve the goodness of the fit (F-test probability = $0.1$ compared to the fit where $\theta$ is allowed to vary) (see Table \ref{reflionx-beta-zxipcf} and Fig.\ref{Mrk841_ldata}, middle panel).   
  
  \setcounter{figure}{8}
   \begin{figure}[pt]
	 \centering
   \resizebox{\hsize}{!}{\includegraphics[height=5cm]{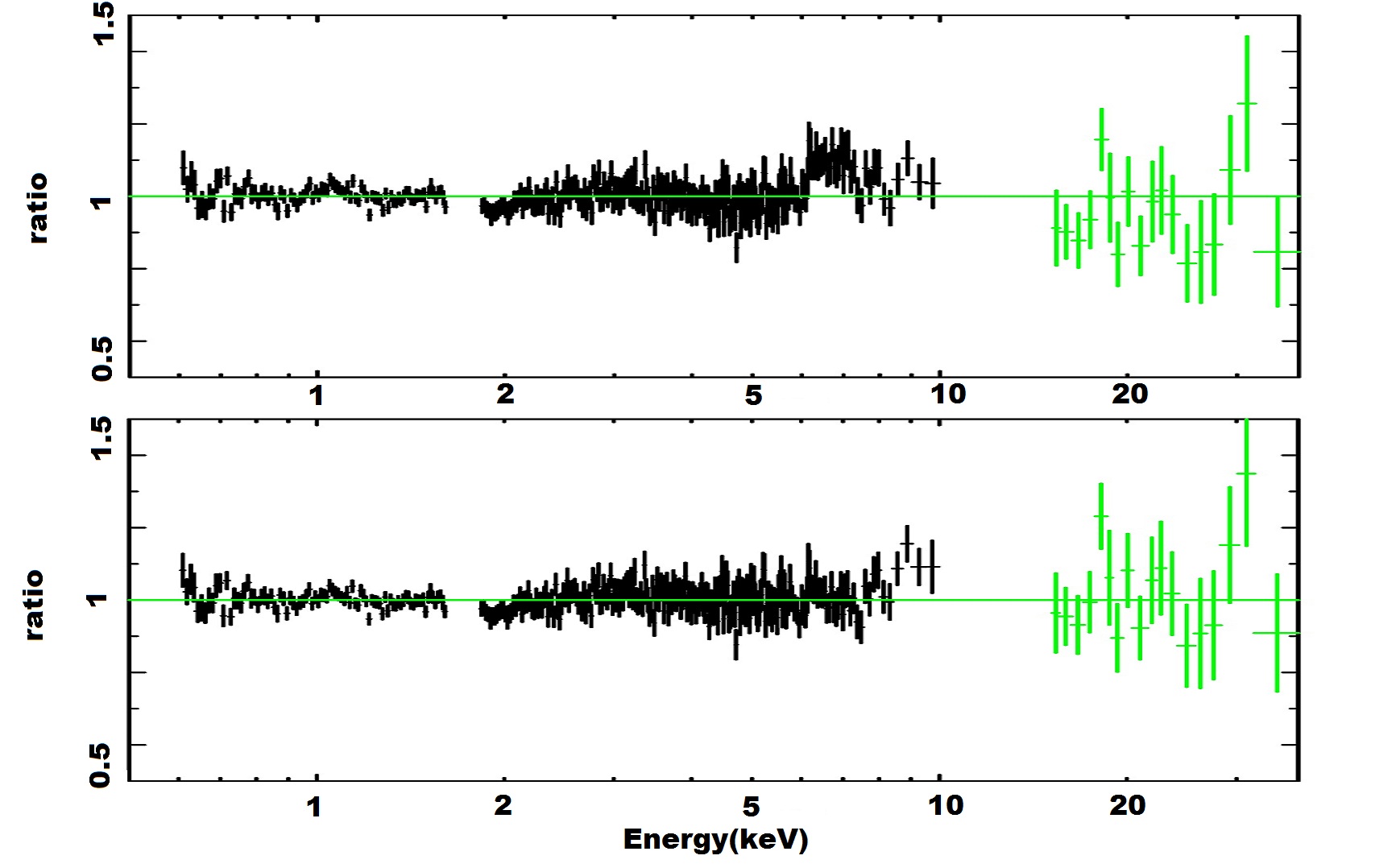}}
     \caption{Ratio of the Mrk~509 data to the blurred reflection model either with (bottom) or without (top) a broad emission K$\alpha$ line (see section 3.6). \textit{XIS/FI} data are in black, while \textit{HXD/PIN} data are in green. For clarity, the data were rebinned, and \textit{XIS1} data are not shown in this plot.}
     \label{reflionratio}
   \end{figure}

 \subsection{Smeared absorption model}
   
  An alternative model to explain the soft excess is a relativistically smeared absorption, where high velocities can broaden all atomic features, producing a curved continuum spectrum. This absorption might correspond to an outflowing disc wind, as shown by \citet{Murray97}. We used \textit{swind} (GD04), one of the \textit{XSpec} models, to describe this component: a \textit{XSTAR} \citep{Bautista01} grid (calculated assuming $\Gamma=2.4$ and a microturbulent velocity of $100\ \textrm{km}/\textrm{s}$) is convolved with a Gaussian function of width $\sigma = v/c$. Free parameters of \textit{swind} model are the column density $N_H$, the ionisation parameter $\xi$ and the velocity dispersion $\sigma$. As for the ionised reflection model, we added a neutral reflection (\textit{reflext}), a narrow Gaussian line for the Fe $K\beta$ emission line and a two-phase warm absorber (\textit{zxipcf} model). As no broad Fe emission line is included in the model, we also added a broad emission line with parameters fixed as in Table \ref{po} (for the \textit{XIS} + \textit{PIN} fit). Best-fit model results are shown in Table \ref{swind1_zxipcf} and Figs. \ref{Mrk509_ldata} and \ref{Mrk841_ldata}, right panel.\\
   For both sources, the values of $\sigma$ are always pegged at the highest authorized value ($v = 0.5\ c$), and we can only give lower limits. 
   It is interesting to cite the results of a fitting without \textit{zxipcf}: $\sigma$ decreases to $\approx0.3$ (it is equal to $0.29^{+0.01}_{-0.02}$ and $0.28^{+0.03}_{-0.02}$ for Mrk~509 and Mrk~841, respectively). We can compare our results for Mrk~841 with P07, who tested the same model for \textit{XMM-Newton} observations: their best-fit values for \textit{swind} model are $2.4\leq \textrm{log}\ \xi \leq3.5$, $0.9 \leq N_H \leq 4.4 \cdot 10^{23}\textrm{cm}^{-2}$, and $\sigma$ consistent with $0.3$ ( but $>0.4$ in two cases). The values of $N_H$ and $\xi$ that we found agree with these results. This model was studied by \citet{Middleton07} (hereafter M07), who analysed the same AGN sample used by C06. They found that, in half of their sample, $\sigma$ is pegged at $0.5$, and that, in general, it has a typical value $>0.3$. Their $\xi$ and $N_H$ best-fit values do not have large dispersion: log~$\xi$ varies between $2.5$ and $3.7$, with lower values (log~$\xi<3$) for their narrow-line Seyfert 1 (NLS1) galaxy sample; the values of $N_H$ are between $5\cdot10^{22}$ and $5\cdot10^{24}$, with, once again, values lower than $1.5\cdot 10^{23}$ restricted to NLS1 galaxies. The intrinsic power-law spectral index varies between $2.0$ and $3.4$, with a clear difference between the softer NLS1s and the harder broad-line AGN. The values that we found agree with these global results, but $\Gamma$ is only marginally compatible with M07's results. The evaluation of $\Gamma$ is important since it affects the soft excess strength and shape, and therefore the \textit{swind} parameters. To examine whether \textit{PIN} data have an effect on the evaluation of $\Gamma$, we tried to fit \textit{XIS} data alone, but the best-fit values of $\Gamma$ are compatible with the broad-band fit.\\
   
   \section{Discussion and conclusions}
Both Mrk~509 and Mrk~841 are bright Seyfert 1 galaxies that are well-known for their soft X-ray excess, their iron $K\alpha$ emission line, and their spectral variability. We have presented our analysis of new \textit{Suzaku} observations: four of Mrk~509 (from April, 2006 to November, 2006), and two of Mrk~841 (January and July, 2007).
The soft excess has been found to be clearly present in both sources, with a flux $\approx2$ times higher than the $3$-$10$ keV power-law extrapolation at $0.5$ keV. \citet{Pounds02} analysed a sample of six Seyfert 1 galaxies, showing that a variety of soft excess shapes exists, correlated with AGN luminosity: our sources during \textit{Suzaku} observations can be described as 'gradual soft excess' galaxies, according to Pounds and Reeves terminology. We can compare our results to past observations (\citealp{Pounds01,Pounds02,DeRosa04}; P07): a soft excess two times higher than the power-law extrapolation at $0.5$ keV is common for both sources, even though Mrk~841 during \textit{XMM-Newton} observations showed a stronger soft excess, $2.5$-$3$ times higher than the power law, and the soft excess of Mrk~509 during obs.1 is quite low, similar to what was observed during the highest state in \textit{BeppoSAX} observations \citep{DeRosa04}.\\

Comparing our observations to statistical studies of the soft excess (GD04, C06, M07), we have confirmed that both sources were detected during a weak soft-excess state, setting them always in the lower part of the sample: for the double Comptonisation model, the ''soft excess strength'' ($R_{GD}$, as defined by GD04) of Mrk~509 and Mrk~841 during our observations is between $0.23$ and $0.28$, while the GD04 sample has a typical value of $0.4$-$0.6$; in the reflection model, we found a $R$ parameter between $0.15$ and $0.18$, which is quite low relative to the reflected flux fractions of C06.\\

Both sources exhibit a hard excess at energies greater than $10$ keV. This excess is weak compared to what was observed by \textit{Suzaku} in three other Seyfert 1 galaxies \citep{Walton10}, though they are rather extreme exemples of hard excesses.\\

A narrow Fe $K\alpha$ emission line is detected in both sources, consistent with a reflection by remote and cold material (i.e. the dust torus or the furthest regions of the accretion disc). For Mrk~841, we compared our results to previous \textit{XMM-Newton} observations, showing that the line flux is constant in time. For an analysis of the Fe K complex emission of Mrk~509, we refer to \citet{Ponti09}. A broad Fe K emission line is observed as well, with equivalent width $\approx 150$ eV in both sources.\\

We detected a two-phase warm absorber in Mrk~509 data, with log~$\xi_{WA,1} \approx 1.7$ and log~$\xi_{WA,2} \approx 2.7$, values that are lower than those found by Smith et al. and Detmers et al.. In the \textit{Suzaku} data, we did not detect the low ionisation phase (log~$\xi = 0.6-0.9$) of the absorber observed by \textit{XMM/RGS}. For Mrk~841, we found a low-ionisation warm absorber with log~$\xi_{WA,1} \approx 2.2$, consistent with the lower ionisation phase in the two-phase absorber described by \citet{Longinotti10}, and we fixed the high ionisation parameter to be  log~$\xi_{WA,2} = 3.3$. For Mrk~509, considering a warm absorber covering factor lower than $1$, always improves the fit.\\

We started the analysis of the broad-band spectra of the two sources by considering the double Comptonisation model, which correctly fits the data ($\tilde{\chi^2} \approx 1.09$ for Mrk~509, and $\tilde{\chi^2} \approx 1.00$ for Mrk~841), but the constancy of the temperature of the cold Comptonisation region in statistical studies of Seyfert 1 galaxies is not explained by current physical models of accretion disc atmospheres, and thus disfavours this model.\\  
We then fitted the broad-band spectrum with the two competitive models developed to explain the soft excess with atomic processes: a blurred reflection and a smeared ionised absorption.\\
For Mrk~509, we need to add a broad emission line to correctly fit the data, and we obtain $\tilde{\chi^2} \approx 1.04$-$1.02$ for the blurred reflection and the smeared absorption model, respectively.\\
The physical origin of this emission is unclear: for the reflection model, a broadened emission line is already included in the code we used, and it is difficult to explain why we do need some more reflection at $E\approx6.4$ keV. Interestingly, this result suggests that the reflection model correctly describes the soft excess shape and strength, but underestimates the broad emission at Fe K complex energies.\\
For the smeared absorption model, we did not include any reflection from the accretion disc, and the fact that we need to add a broad emission line is less critical: the \textit{swind} code assumes a simple power-law emission as input for the smeared absorption, and the presence of residuals in the Fe K complex region means that the continuum spectrum absorbed by the wind does include, indeed, some reflected component.\\
For Mrk~841, no extra emission in the reflection model is required, while a broad Gaussian emission line is needed in the absorption model. Each model gives a good fit to the data, and none of them can be invalidated ($\tilde{\chi^2} \approx 1.04$ for the two models).\\

\textit{Suzaku} data, characterised by a good energy resolution in a large energy domain ($0.5$-$60$ keV), seemed to be able to constrain the intrinsic power-law photon index at high energies, and then to favor one model over another. The best-fit models show that, while in the soft X-ray domain the models are degenerate, in the hard energy band they do not superpose any more and we could, in principle, distinguish between them, but the \textit{PIN} signal-to-noise ratio is still too low to constrain the fit. For the sum of Mrk~509 observations, we can estimate from the best-fit model in the hard X-ray domain a
flux $F_{17-60keV} = 4.47$, $4.81$, and $4.27\cdot 10^{-11}\ \textrm{ergs}\ \textrm{cm}^{-2}\ \textrm{s}^{-1}$ and $F_{14-150keV} = 7.64$, $8.07$, and $7.66\cdot 10^{-11}\ \textrm{ergs}\ \textrm{cm}^{-2}\ \textrm{s}^{-1}$ for the double Comptonisation, the blurred reflection, and the smeared absorption model, respectively; for the sum of Mrk~841 observations we obtain $F_{17-60keV} = 1.94$, $2.15$, and $1.67\cdot 10^{-11}\ \textrm{ergs}\ \textrm{cm}^{-2}\ \textrm{s}^{-1}$ and $F_{14-150keV} = 3.47$, $3.72$, and $3.06\cdot 10^{-11}\ \textrm{ergs}\ \textrm{cm}^{-2}\ \textrm{s}^{-1}$ for the three models, respectively. These values can be compared with the \textit{INTEGRAL} \citep{Krivonos10} and \textit{Swift/BAT} \citep{Cusumano10} detections of our sources: \textit{INTEGRAL} detected Mrk~509 at $F_{17-60keV} = 5.90\cdot 10^{-11}\ \textrm{ergs}\ \textrm{cm}^{-2}\ \textrm{s}^{-1}$ and Mrk~841 at $F_{17-60keV} = 3.07\cdot 10^{-11}\ \textrm{ergs}\ \textrm{cm}^{-2}\ \textrm{s}^{-1}$; \textit{Swift/BAT} observed a flux in the $14$-$150$ keV energy band equal to $8.5\cdot 10^{-11}\ \textrm{ergs}\ \textrm{cm}^{-2}\ \textrm{s}^{-1}$ for Mrk~509 and $2.9\cdot 10^{-11}\ \textrm{ergs}\ \textrm{cm}^{-2}\ \textrm{s}^{-1}$ for Mrk~841.  
The differences between the predicted high-energy fluxes demonstrate that a simultaneous broad-band observation with adequate signal-to-noise ratio would be able to distinguish between the different models.\\ Statistical studies of AGN can help us to solve the problem of the origin of the soft excess. The discovery of the remarkably constant temperature of the cold Comptonisation region, despite the AGN luminosity and black hole mass spanning several decades, is hardly explained in a double Comptonisation scenario. Our $kT_s$ values agree with GD04 results, but we note that adding a warm absorber (which was not included in the GD04 analysis) affects the $kT_s$ best-fit values.\\
Reflection and absorption models, that are tied to atomic processes, can easily explain the small spread in the soft-excess temperature. 
However, the smeared absorption model has the weakness that the velocity dispersion required to fit the soft excess is very large, reaching $c/2$, which corresponds to extremely relativistic terminal velocities of the wind. Numerical simulations \citep{Schurch08} show that disc winds have $\sigma<0.1c$, which is very low compared to \textit{swind} model results. An outlet could be a magnetically driven wind, partially covering the source. We underline that \textit{swind} takes into account only the absorption by the wind, but if we also consider its emission, \citep{Chevallier06, Schurch06} the fit requires a lower smearing velocity and a lower column density. A possible concern of the \textit{swind} model is that the absorption is evaluated assuming an incident power-law spectrum with $\Gamma=2.4$, which is inconsistent with the $\Gamma$ best-fit values of our observations.\\
The Fe broad-line excess we observe in both sources suggests that, even in an absorption scenario, we need some reflected emission from the inner regions of the accretion disc. A broad Fe K emission line can also be reproduced by disk wind models \citep{Sim10}, with lower outflow velocities ($0.1$c).\\
The blurred reflection model fits the \textit{Suzaku} data well for Mrk~841, while for Mrk~509 we need to include an additional emission at Fe K energies, which is difficult to explain. The model does not need a high reflection fraction R, which is seen in several other cases, and is explained by a huge light bending effect \citep{MiniuttiFabian} close to the black hole.\\
It is of course possible that different processes are at work at the same time. In this scenario (which is difficult to test in a stationary framework owing to the high number of free parameters involved), the broad emission observed in Mrk~509 could be due to standard ionised reflection from the accretion disc, while Comptonisation or smeared absorption should be responsible for the soft excess below $1$ keV (see \citet{Patrick11}).\\
Alternatively, the continuum X-ray emission from the hot corona might deviate from a power law when measured over a broad energy band, showing an intrinsic curvature that has not been taken into account in our modeling.\\
Variability studies can help us ascertain the physical origin of the soft excess. In the blurred reflection scenario, variability can be explained by changes in the height of the corona above the disc, but this picture implies a strong continuum-reflected component correlation that has not been observed. In the smeared absorption model, spectral variability is more easily explained by a variation in the absorber physical parameters, or its covering factor. Unluckily, neither of our sources displayed significant variability during \textit{Suzaku} observations, and we could not use temporal information to more tightly constrain our models. \\ 
Future X-ray observatories, and in particular \textit{Astro-H} \citep{AstroH} and \textit{NuSTAR} \citep{NuSTAR}, wich have higher sensitivities and energy resolutions in the hard X-ray domain, will be able to more tightly constrain the spectrum, casting light on the physical origin of the soft excess in Seyfert 1 galaxies. \\

\section{Acknowledgments}

This research has made use of data obtained from the Suzaku satellite, a collaborative mission between the space agencies of Japan (JAXA) and the USA (NASA). This work was financially supported by the GdR PCHE in France. MC thanks Matteo Guainazzi for useful discussions and for his help in rebinning the data. GP acknowledges support via an EU Marie Curie Intra-European
Fellowship under contract no. FP7-PEOPLE-2009-IEF-254279. POP acknowledges financial support from the CNES agency. We also thank the anonymous referee for providing constructive remarks that improved the quality of the paper. 

   \bibliographystyle{aa}
   \bibliography{Mrks_Suzaku}

\begin{thebibliography}{70}
\expandafter\ifx\csname natexlab\endcsname\relax\def\natexlab#1{#1}\fi

\bibitem[{{Abdo} {et~al.}(2010){Abdo}, {Ackermann}, {Ajello}, {Atwood},
  {Axelsson}, {Baldini}, {Ballet}, {Barbiellini}, {Bastieri}, {Baughman},
  {Bechtol}, {Bellazzini}, {Berenji}, {Bloom}, {Bogaert}, {Bonamente},
  {Borgland}, {Bregeon}, {Brez}, {Brigida}, {Bruel}, {Burnett}, {Caliandro},
  {Cameron}, {Caraveo}, {Casandjian}, {Cavazzuti}, {Cecchi}, {{\c C}elik},
  {Chekhtman}, {Cheung}, {Chiang}, {Ciprini}, {Claus}, {Cohen-Tanugi},
  {Conrad}, {Cutini}, {Dermer}, {de Angelis}, {de Palma}, {Digel}, {Silva},
  {Drell}, {Dubois}, {Dumora}, {Farnier}, {Favuzzi}, {Fegan}, {Ferrara},
  {Focke}, {Frailis}, {Fuhrmann}, {Fukazawa}, {Funk}, {Fusco}, {Gargano},
  {Gasparrini}, {Gehrels}, {Germani}, {Giebels}, {Giglietto}, {Giordano},
  {Giroletti}, {Glanzman}, {Godfrey}, {Grenier}, {Grondin}, {Grove},
  {Guillemot}, {Guiriec}, {Hanabata}, {Harding}, {Hayashida}, {Hays}, {Hughes},
  {J{\'o}hannesson}, {Johnson}, {Johnson}, {Johnson}, {Kadler}, {Kamae},
  {Katagiri}, {Kataoka}, {Kerr}, {Kn{\"o}dlseder}, {Kocian}, {Kuehn}, {Kuss},
  {Lande}, {Latronico}, {Lemoine-Goumard}, {Longo}, {Loparco}, {Lott},
  {Lovellette}, {Lubrano}, {Madejski}, {Makeev}, {Marelli}, {Massaro},
  {Max-Moerbeck}, {Mazziotta}, {McConville}, {McEnery}, {Meurer}, {Michelson},
  {Mitthumsiri}, {Mizuno}, {Moiseev}, {Monte}, {Monzani}, {Morselli},
  {Moskalenko}, {Murgia}, {Nolan}, {Norris}, {Nuss}, {Ohsugi}, {Omodei},
  {Orlando}, {Ormes}, {Ozaki}, {Paneque}, {Panetta}, {Parent}, {Pavlidou},
  {Pearson}, {Pelassa}, {Pepe}, {Pesce-Rollins}, {Piron}, {Porter},
  {Rain{\`o}}, {Rando}, {Razzano}, {Razzaque}, {Readhead}, {Reimer}, {Reimer},
  {Reposeur}, {Richards}, {Ritz}, {Rochester}, {Rodriguez}, {Romani}, {Roth},
  {Ryde}, {Sadrozinski}, {Sanchez}, {Sander}, {Saz Parkinson}, {Scargle},
  {Sgr{\`o}}, {Shaw}, {Siskind}, {Smith}, {Smith}, {Spandre}, {Spinelli},
  {Stevenson}, {Strickman}, {Suson}, {Tajima}, {Takahashi}, {Tanaka}, {Thayer},
  {Thayer}, {Thompson}, {Tibaldo}, {Tibolla}, {Torres}, {Tosti}, {Tramacere},
  {Ubertini}, {Uchiyama}, {Usher}, {Vasileiou}, {Vilchez}, {Vitale}, {Waite},
  {Wang}, {Winer}, {Wood}, {Yasuda}, {Ylinen}, {Zensus}, {Ziegler}, {The FERMI
  LAT Collaboration}, {Angelakis}, {Hovatta}, {Hoversten}, {Ikejiri},
  {Kawabata}, {Kovalev}, {Kovalev}, {Krichbaum}, {Lister},
  {L{\"a}hteenm{\"a}ki}, {Marchili}, {Ogle}, {Pagani}, {Pushkarev}, {Sakimoto},
  {Sasada}, {Tornikoski}, {Uemura}, {Yamanaka}, \& {Yamashita}}]{Abdo10}
{Abdo}, A.~A., {Ackermann}, M., {Ajello}, M., {et~al.} 2010, \apj, 710, 810

\bibitem[{{Akiyama} {et~al.}(2003){Akiyama}, {Ueda}, {Ohta}, {Takahashi}, \&
  {Yamada}}]{Akiyama03}
{Akiyama}, M., {Ueda}, Y., {Ohta}, K., {Takahashi}, T., \& {Yamada}, T. 2003,
  \apjs, 148, 275

\bibitem[{{Arnaud} {et~al.}(1985){Arnaud}, {Branduardi-Raymont}, {Culhane},
  {Fabian}, {Hazard}, {McGlynn}, {Shafer}, {Tennant}, \& {Ward}}]{Arnaud85}
{Arnaud}, K.~A., {Branduardi-Raymont}, G., {Culhane}, J.~L., {et~al.} 1985,
  \mnras, 217, 105

\bibitem[{{Bautista} \& {Kallman}(2001)}]{Bautista01}
{Bautista}, M.~A. \& {Kallman}, T.~R. 2001, \apjs, 134, 139

\bibitem[{{Bianchi} {et~al.}(2001){Bianchi}, {Matt}, {Haardt}, {Maraschi},
  {Nicastro}, {Perola}, {Petrucci}, \& {Piro}}]{Bianchi01}
{Bianchi}, S., {Matt}, G., {Haardt}, F., {et~al.} 2001, \aap, 376, 77

\bibitem[{{Boldt} \& {Leiter}(1987)}]{CXB}
{Boldt}, E. \& {Leiter}, D. 1987, \apjl, 322, L1

\bibitem[{{Chevallier} {et~al.}(2006){Chevallier}, {Collin}, {Dumont},
  {Czerny}, {Mouchet}, {Gon{\c c}alves}, \& {Goosmann}}]{Chevallier06}
{Chevallier}, L., {Collin}, S., {Dumont}, A., {et~al.} 2006, \aap, 449, 493

\bibitem[{{Cooke} {et~al.}(1978){Cooke}, {Ricketts}, {Maccacaro}, {Pye},
  {Elvis}, {Watson}, {Griffiths}, {Pounds}, {McHardy}, {Maccagni}, {Seward},
  {Page}, \& {Turner}}]{Cooke78}
{Cooke}, B.~A., {Ricketts}, M.~J., {Maccacaro}, T., {et~al.} 1978, \mnras, 182,
  489

\bibitem[{{Crummy} {et~al.}(2006){Crummy}, {Fabian}, {Gallo}, \&
  {Ross}}]{Crummy06}
{Crummy}, J., {Fabian}, A.~C., {Gallo}, L., \& {Ross}, R.~R. 2006, \mnras, 365,
  1067

\bibitem[{{Cusumano} {et~al.}(2010){Cusumano}, {La Parola}, {Segreto},
  {Ferrigno}, {Maselli}, {Sbarufatti}, {Romano}, {Chincarini}, {Giommi},
  {Masetti}, {Moretti}, {Parisi}, \& {Tagliaferri}}]{Cusumano10}
{Cusumano}, G., {La Parola}, V., {Segreto}, A., {et~al.} 2010, \aap, 524, A64+

\bibitem[{{Czerny} {et~al.}(2003){Czerny}, {Niko{\l}ajuk},
  {R{\'o}{\.z}a{\'n}ska}, {Dumont}, {Loska}, \& {Zycki}}]{Czerny03}
{Czerny}, B., {Niko{\l}ajuk}, M., {R{\'o}{\.z}a{\'n}ska}, A., {et~al.} 2003,
  \aap, 412, 317

\bibitem[{{Dadina} {et~al.}(2005){Dadina}, {Cappi}, {Malaguti}, {Ponti}, \& {de
  Rosa}}]{Dadina05}
{Dadina}, M., {Cappi}, M., {Malaguti}, G., {Ponti}, G., \& {de Rosa}, A. 2005,
  \aap, 442, 461

\bibitem[{{de La Calle P{\'e}rez} {et~al.}(2010){de La Calle P{\'e}rez},
  {Longinotti}, {Guainazzi}, {Bianchi}, {Dov{\v c}iak}, {Cappi}, {Matt},
  {Miniutti}, {Petrucci}, {Piconcelli}, {Ponti}, {Porquet}, \&
  {Santos-Lle{\'o}}}]{delaCalle10}
{de La Calle P{\'e}rez}, I., {Longinotti}, A.~L., {Guainazzi}, M., {et~al.}
  2010, \aap, 524, A50+

\bibitem[{{De Rosa} {et~al.}(2004){De Rosa}, {Piro}, {Matt}, \&
  {Perola}}]{DeRosa04}
{De Rosa}, A., {Piro}, L., {Matt}, G., \& {Perola}, G.~C. 2004, \aap, 413, 895

\bibitem[{{Detmers} {et~al.}(2010){Detmers}, {Kaastra}, {Costantini},
  {Verbunt}, {Cappi}, \& {de Vries}}]{Detmers10}
{Detmers}, R.~G., {Kaastra}, J.~S., {Costantini}, E., {et~al.} 2010, \aap, 516,
  A61+

\bibitem[{{Dickey} \& {Lockman}(1990)}]{Dickey90}
{Dickey}, J.~M. \& {Lockman}, F.~J. 1990, \araa, 28, 215

\bibitem[{{Done} {et~al.}(2007){Done}, {Gierli{\'n}ski}, \& {Kubota}}]{DoneCYG}
{Done}, C., {Gierli{\'n}ski}, M., \& {Kubota}, A. 2007, \aapr, 15, 1

\bibitem[{{Ebrero} {et~al.}(2010){Ebrero}, {Costantini}, {Kaastra}, {Detmers},
  {Arav}, {Kriss}, {Korista}, \& {Steenbrugge}}]{Ebrero2010}
{Ebrero}, J., {Costantini}, E., {Kaastra}, J.~S., {et~al.} 2010, \aap, 520,
  A36+

\bibitem[{{Fisher} {et~al.}(1995){Fisher}, {Huchra}, {Strauss}, {Davis},
  {Yahil}, \& {Schlegel}}]{Fisher95}
{Fisher}, K.~B., {Huchra}, J.~P., {Strauss}, M.~A., {et~al.} 1995, \apjs, 100,
  69

\bibitem[{{George} {et~al.}(1993){George}, {Nandra}, {Fabian}, {Turner},
  {Done}, \& {Day}}]{George93}
{George}, I.~M., {Nandra}, K., {Fabian}, A.~C., {et~al.} 1993, \mnras, 260, 111

\bibitem[{{George} {et~al.}(1994){George}, {Nandra}, {Turner}, \&
  {Celotti}}]{George94}
{George}, I.~M., {Nandra}, K., {Turner}, T.~J., \& {Celotti}, A. 1994, \apjl,
  436, L59

\bibitem[{{Gierli{\'n}ski} \& {Done}(2004)}]{Gierlinski04}
{Gierli{\'n}ski}, M. \& {Done}, C. 2004, \mnras, 349, L7

\bibitem[{{Haardt} \& {Maraschi}(1993)}]{Haardt93}
{Haardt}, F. \& {Maraschi}, L. 1993, \apj, 413, 507

\bibitem[{{Harrison} {et~al.}(2010){Harrison}, {Boggs}, {Christensen}, {Craig},
  {Hailey}, {Stern}, {Zhang}, {Angelini}, {An}, {Bhalereo}, {Brejnholt},
  {Cominsky}, {Cook}, {Doll}, {Giommi}, {Grefenstette}, {Hornstrup}, {Kaspi},
  {Kim}, {Kitaguchi}, {Koglin}, {Liebe}, {Madejski}, {Kruse Madsen}, {Mao},
  {Meier}, {Miyasaka}, {Mori}, {Perri}, {Pivovaroff}, {Puccetti}, {Rana}, \&
  {Zoglauer}}]{NuSTAR}
{Harrison}, F.~A., {Boggs}, S., {Christensen}, F., {et~al.} 2010, in Society of
  Photo-Optical Instrumentation Engineers (SPIE) Conference Series, Vol. 7732,
  Society of Photo-Optical Instrumentation Engineers (SPIE) Conference Series

\bibitem[{{Janiuk} {et~al.}(2001){Janiuk}, {Czerny}, \& {Madejski}}]{Janiuk01}
{Janiuk}, A., {Czerny}, B., \& {Madejski}, G.~M. 2001, \apj, 557, 408

\bibitem[{{Kaastra} {et~al.}(2011){Kaastra}, {Petrucci}, {Cappi}, {Arav},
  {Behar}, {Bianchi}, {Bloom}, {Blustin}, {Branduardi-Raymont}, {Costantini},
  {Dadina}, {Detmers}, {Ebrero}, {Jonker}, {Klein}, {Kriss}, {Lubi{\'n}ski},
  {Malzac}, {Mehdipour}, {Paltani}, {Pinto}, {Ponti}, {Ratti}, {Smith},
  {Steenbrugge}, \& {de Vries}}]{Kaastra11}
{Kaastra}, J.~S., {Petrucci}, P.-O., {Cappi}, M., {et~al.} 2011, \aap, 534, A36

\bibitem[{{Koyama} {et~al.}(2007){Koyama}, {Tsunemi}, {Dotani}, {Bautz},
  {Hayashida}, {Tsuru}, {Matsumoto}, {Ogawara}, {Ricker}, {Doty}, {Kissel},
  {Foster}, {Nakajima}, {Yamaguchi}, {Mori}, {Sakano}, {Hamaguchi},
  {Nishiuchi}, {Miyata}, {Torii}, {Namiki}, {Katsuda}, {Matsuura}, {Miyauchi},
  {Anabuki}, {Tawa}, {Ozaki}, {Murakami}, {Maeda}, {Ichikawa}, {Prigozhin},
  {Boughan}, {Lamarr}, {Miller}, {Burke}, {Gregory}, {Pillsbury}, {Bamba},
  {Hiraga}, {Senda}, {Katayama}, {Kitamoto}, {Tsujimoto}, {Kohmura}, {Tsuboi},
  \& {Awaki}}]{XIS}
{Koyama}, K., {Tsunemi}, H., {Dotani}, T., {et~al.} 2007, \pasj, 59, 23

\bibitem[{{Krivonos} {et~al.}(2010){Krivonos}, {Tsygankov}, {Revnivtsev},
  {Grebenev}, {Churazov}, \& {Sunyaev}}]{Krivonos10}
{Krivonos}, R., {Tsygankov}, S., {Revnivtsev}, M., {et~al.} 2010, \aap, 523,
  A61+

\bibitem[{{Laor}(1991)}]{Laor91}
{Laor}, A. 1991, \apj, 376, 90

\bibitem[{{Longinotti} {et~al.}(2010){Longinotti}, {Costantini}, {Petrucci},
  {Boisson}, {Mouchet}, {Santos-Lleo}, {Matt}, {Ponti}, \& {Gon{\c
  c}alves}}]{Longinotti10}
{Longinotti}, A.~L., {Costantini}, E., {Petrucci}, P.~O., {et~al.} 2010, \aap,
  510, A92+

\bibitem[{{Magdziarz} {et~al.}(1998){Magdziarz}, {Blaes}, {Zdziarski},
  {Johnson}, \& {Smith}}]{Magdziarz98}
{Magdziarz}, P., {Blaes}, O.~M., {Zdziarski}, A.~A., {Johnson}, W.~N., \&
  {Smith}, D.~A. 1998, \mnras, 301, 179

\bibitem[{{Magdziarz} \& {Zdziarski}(1995)}]{pexrav}
{Magdziarz}, P. \& {Zdziarski}, A.~A. 1995, \mnras, 273, 837

\bibitem[{{Middleton} {et~al.}(2007){Middleton}, {Done}, \&
  {Gierli{\'n}ski}}]{Middleton07}
{Middleton}, M., {Done}, C., \& {Gierli{\'n}ski}, M. 2007, \mnras, 381, 1426

\bibitem[{{Miniutti} \& {Fabian}(2004)}]{MiniuttiFabian}
{Miniutti}, G. \& {Fabian}, A.~C. 2004, \mnras, 349, 1435

\bibitem[{{Miniutti} {et~al.}(2009){Miniutti}, {Ponti}, {Greene}, {Ho},
  {Fabian}, \& {Iwasawa}}]{Miniutti09}
{Miniutti}, G., {Ponti}, G., {Greene}, J.~E., {et~al.} 2009, \mnras, 394, 443

\bibitem[{{Miyazawa} {et~al.}(2009){Miyazawa}, {Haba}, \&
  {Kunieda}}]{Miyazawa09}
{Miyazawa}, T., {Haba}, Y., \& {Kunieda}, H. 2009, \pasj, 61, 1331

\bibitem[{{Morini} {et~al.}(1987){Morini}, {Lipani}, \& {Molteni}}]{Morini87}
{Morini}, M., {Lipani}, N.~A., \& {Molteni}, D. 1987, \apj, 317, 145

\bibitem[{{Murray} \& {Chiang}(1997)}]{Murray97}
{Murray}, N. \& {Chiang}, J. 1997, \apj, 474, 91

\bibitem[{{Palmeri} {et~al.}(2003){Palmeri}, {Mendoza}, {Kallman}, {Bautista},
  \& {Mel{\'e}ndez}}]{Palmeri03}
{Palmeri}, P., {Mendoza}, C., {Kallman}, T.~R., {Bautista}, M.~A., \&
  {Mel{\'e}ndez}, M. 2003, \aap, 410, 359

\bibitem[{{Patrick} {et~al.}(2011){Patrick}, {Reeves}, {Porquet}, {Markowitz},
  {Lobban}, \& {Terashima}}]{Patrick11}
{Patrick}, A.~R., {Reeves}, J.~N., {Porquet}, D., {et~al.} 2011, \mnras, 411,
  2353

\bibitem[{{Perola} {et~al.}(2000){Perola}, {Matt}, {Fiore}, {Grandi},
  {Guainazzi}, {Haardt}, {Maraschi}, {Mineo}, {Nicastro}, \& {Piro}}]{Perola00}
{Perola}, G.~C., {Matt}, G., {Fiore}, F., {et~al.} 2000, \aap, 358, 117

\bibitem[{{Petrucci} {et~al.}(2007){Petrucci}, {Ponti}, {Matt}, {Longinotti},
  {Malzac}, {Mouchet}, {Boisson}, {Maraschi}, {Nandra}, \&
  {Ferrando}}]{Petrucci07}
{Petrucci}, P.~O., {Ponti}, G., {Matt}, G., {et~al.} 2007, \aap, 470, 889

\bibitem[{{Piconcelli} {et~al.}(2005){Piconcelli}, {Jimenez-Bail{\'o}n},
  {Guainazzi}, {Schartel}, {Rodr{\'{\i}}guez-Pascual}, \&
  {Santos-Lle{\'o}}}]{Piconcelli05}
{Piconcelli}, E., {Jimenez-Bail{\'o}n}, E., {Guainazzi}, M., {et~al.} 2005,
  \aap, 432, 15

\bibitem[{{Ponti} {et~al.}(2009){Ponti}, {Cappi}, {Vignali}, {Miniutti},
  {Tombesi}, {Dadina}, {Fabian}, {Grandi}, {Kaastra}, {Petrucci}, {Bianchi},
  {Matt}, {Maraschi}, \& {Malaguti}}]{Ponti09}
{Ponti}, G., {Cappi}, M., {Vignali}, C., {et~al.} 2009, \mnras, 394, 1487

\bibitem[{{Ponti} {et~al.}(2006){Ponti}, {Miniutti}, {Cappi}, {Maraschi},
  {Fabian}, \& {Iwasawa}}]{Ponti06}
{Ponti}, G., {Miniutti}, G., {Cappi}, M., {et~al.} 2006, \mnras, 368, 903

\bibitem[{{Pounds} \& {Reeves}(2002)}]{Pounds02}
{Pounds}, K. \& {Reeves}, J. 2002, ArXiv:astro-ph/0201436, Proceedings of the
  Symposium on `New Visions of the X-ray Universe in the XMM-Newton and Chandra
  Era', 26-30 November 2001, ESTEC, The Netherlands

\bibitem[{{Pounds} {et~al.}(2001){Pounds}, {Reeves}, {O'Brien}, {Page},
  {Turner}, \& {Nayakshin}}]{Pounds01}
{Pounds}, K., {Reeves}, J., {O'Brien}, P., {et~al.} 2001, \apj, 559, 181

\bibitem[{{Pounds} {et~al.}(1994){Pounds}, {Nandra}, {Fink}, \&
  {Makino}}]{Pounds94}
{Pounds}, K.~A., {Nandra}, K., {Fink}, H.~H., \& {Makino}, F. 1994, \mnras,
  267, 193

\bibitem[{{Pravdo} {et~al.}(1981){Pravdo}, {Nugent}, {Nousek}, {Jensen},
  {Wilson}, \& {Becker}}]{Pravdo}
{Pravdo}, S., {Nugent}, J., {Nousek}, J., {et~al.} 1981, \apj, 251, 501

\bibitem[{{Reeves} {et~al.}(2008){Reeves}, {Done}, {Pounds}, {Terashima},
  {Hayashida}, {Anabuki}, {Uchino}, \& {Turner}}]{Reeves08}
{Reeves}, J., {Done}, C., {Pounds}, K., {et~al.} 2008, \mnras, 385, L108

\bibitem[{{Reynolds}(1997)}]{Reynolds97}
{Reynolds}, C.~S. 1997, \mnras, 286, 513

\bibitem[{{Ross} \& {Fabian}(2005)}]{RossFabian}
{Ross}, R.~R. \& {Fabian}, A.~C. 2005, \mnras, 358, 211

\bibitem[{{Ross} {et~al.}(1999){Ross}, {Fabian}, \& {Young}}]{Ross99}
{Ross}, R.~R., {Fabian}, A.~C., \& {Young}, A.~J. 1999, \mnras, 306, 461

\bibitem[{{Sazonov} {et~al.}(2007){Sazonov}, {Revnivtsev}, {Krivonos},
  {Churazov}, \& {Sunyaev}}]{Sazonov07}
{Sazonov}, S., {Revnivtsev}, M., {Krivonos}, R., {Churazov}, E., \& {Sunyaev},
  R. 2007, \aap, 462, 57

\bibitem[{{Schurch} \& {Done}(2006)}]{Schurch06}
{Schurch}, N.~J. \& {Done}, C. 2006, \mnras, 371, 81

\bibitem[{{Schurch} \& {Done}(2008)}]{Schurch08}
{Schurch}, N.~J. \& {Done}, C. 2008, \mnras, 386, L1

\bibitem[{{Shakura} \& {Sunyaev}(1976)}]{Shakura76}
{Shakura}, N.~I. \& {Sunyaev}, R.~A. 1976, \mnras, 175, 613

\bibitem[{{Sim} {et~al.}(2010){Sim}, {Miller}, {Long}, {Turner}, \&
  {Reeves}}]{Sim10}
{Sim}, S.~A., {Miller}, L., {Long}, K.~S., {Turner}, T.~J., \& {Reeves}, J.~N.
  2010, \mnras, 404, 1369

\bibitem[{{Singh} {et~al.}(1985){Singh}, {Garmire}, \& {Nousek}}]{Singh85}
{Singh}, K.~P., {Garmire}, G.~P., \& {Nousek}, J. 1985, \apj, 297, 633

\bibitem[{{Singh} {et~al.}(1990){Singh}, {Westergaard}, {Schnopper}, {Awaki},
  \& {Tawara}}]{Singh90}
{Singh}, K.~P., {Westergaard}, N.~J., {Schnopper}, H.~W., {Awaki}, H., \&
  {Tawara}, Y. 1990, \apj, 363, 131

\bibitem[{{Smith} {et~al.}(1977){Smith}, {Burbidge}, {Baldwin}, {Tohline},
  {Wampler}, {Hazard}, \& {Murdoch}}]{Smith77}
{Smith}, H.~E., {Burbidge}, E.~M., {Baldwin}, J.~A., {et~al.} 1977, \apj, 215,
  427

\bibitem[{{Smith} {et~al.}(2007){Smith}, {Page}, \&
  {Branduardi-Raymont}}]{Smith07}
{Smith}, R.~A.~N., {Page}, M.~J., \& {Branduardi-Raymont}, G. 2007, \aap, 461,
  135

\bibitem[{{Sobolewska} \& {Done}(2007)}]{Sobolewska07}
{Sobolewska}, M.~A. \& {Done}, C. 2007, \mnras, 374, 150

\bibitem[{{Takahashi} {et~al.}(2007){Takahashi}, {Abe}, {Endo}, {Endo}, {Ezoe},
  {Fukazawa}, {Hamaya}, {Hirakuri}, {Hong}, {Horii}, {Inoue}, {Isobe}, {Itoh},
  {Iyomoto}, {Kamae}, {Kasama}, {Kataoka}, {Kato}, {Kawaharada}, {Kawano},
  {Kawashima}, {Kawasoe}, {Kishishita}, {Kitaguchi}, {Kobayashi}, {Kokubun},
  {Kotoku}, {Kouda}, {Kubota}, {Kuroda}, {Madejski}, {Makishima}, {Masukawa},
  {Matsumoto}, {Mitani}, {Miyawaki}, {Mizuno}, {Mori}, {Mori}, {Murashima},
  {Murakami}, {Nakazawa}, {Niko}, {Nomachi}, {Okada}, {Ohno}, {Oonuki}, {Ota},
  {Ozawa}, {Sato}, {Shinoda}, {Sugiho}, {Suzuki}, {Taguchi}, {Takahashi},
  {Takahashi}, {Takeda}, {Tamura}, {Tamura}, {Tanaka}, {Tanihata}, {Tashiro},
  {Terada}, {Tominaga}, {Uchiyama}, {Watanabe}, {Yamaoka}, {Yanagida}, \&
  {Yonetoku}}]{HXD}
{Takahashi}, T., {Abe}, K., {Endo}, M., {et~al.} 2007, \pasj, 59, 35

\bibitem[{{Takahashi} {et~al.}(2010){Takahashi}, {Mitsuda}, {Kelley},
  {Aharonian}, {Akimoto}, {Allen}, {Anabuki}, {Angelini}, {Arnaud}, {Awaki},
  {Bamba}, {Bando}, {Bautz}, {Blandford}, {Boyce}, {Brown}, {Chernyakova},
  {Coppi}, {Costantini}, {Cottam}, {Crow}, {de Plaa}, {de Vries}, {den Herder},
  {Dipirro}, {Done}, {Dotani}, {Ebisawa}, {Enoto}, {Ezoe}, {Fabian},
  {Fujimoto}, {Fukazawa}, {Funk}, {Furuzawa}, {Galeazzi}, {Gandhi}, {Gendreau},
  {Gilmore}, {Haba}, {Hamaguchi}, {Hatsukade}, {Hayashida}, {Hiraga}, {Hirose},
  {Hornschemeier}, {Hughes}, {Hwang}, {Iizuka}, {Ishibashi}, {Ishida},
  {Ishimura}, {Ishisaki}, {Isobe}, {Ito}, {Iwata}, {Kaastra}, {Kallman},
  {Kamae}, {Katagiri}, {Kataoka}, {Katsuda}, {Kawaharada}, {Kawai}, {Kawasaki},
  {Khangaluyan}, {Kilbourne}, {Kinugasa}, {Kitamoto}, {Kitayama}, {Kohmura},
  {Kokubun}, {Kosaka}, {Kotani}, {Koyama}, {Kubota}, {Kunieda}, {Laurent},
  {Lebrun}, {Limousin}, {Loewenstein}, {Long}, {Madejski}, {Maeda},
  {Makishima}, {Markevitch}, {Matsumoto}, {Matsushita}, {McCammon}, {Miller},
  {Mineshige}, {Minesugi}, {Miyazawa}, {Mizuno}, {Mori}, {Mori}, {Mukai},
  {Murakami}, {Murakami}, {Mushotzky}, {Nakagawa}, {Nakagawa}, {Nakajima},
  {Nakamori}, {Nakazawa}, {Namba}, {Nomachi}, {O'Dell}, {Ogawa}, {Ogawa},
  {Ogi}, {Ohashi}, {Ohno}, {Ohta}, {Okajima}, {Ota}, {Ozaki}, {Paerels},
  {Paltani}, {Parmar}, {Petre}, {Pohl}, {Porter}, {Ramsey}, {Reynolds},
  {Sakai}, {Sambruna}, {Sato}, {Sato}, {Serlemitsos}, {Shida}, {Shimada},
  {Shinozaki}, {Shirron}, {Smith}, {Sneiderman}, {Soong}, {Stawarz}, {Sugita},
  {Szymkowiak}, {Tajima}, {Takahashi}, {Takei}, {Tamagawa}, {Tamura}, {Tamura},
  {Tanaka}, {Tanaka}, {Tanaka}, {Tashiro}, {Tawara}, {Terada}, {Terashima},
  {Tombesi}, {Tomida}, {Tozuka}, {Tsuboi}, {Tsujimoto}, {Tsunemi}, {Tsuru},
  {Uchida}, {Uchiyama}, {Uchiyama}, {Ueda}, {Uno}, {Urry}, {Watanabe}, {White},
  {Yamada}, {Yamaguchi}, {Yamaoka}, {Yamasaki}, {Yamauchi}, {Yamauchi},
  {Yatsu}, {Yonetoku}, \& {Yoshida}}]{AstroH}
{Takahashi}, T., {Mitsuda}, K., {Kelley}, R., {et~al.} 2010, in Society of
  Photo-Optical Instrumentation Engineers (SPIE) Conference Series, Vol. 7732,
  Society of Photo-Optical Instrumentation Engineers (SPIE) Conference Series

\bibitem[{{Thorne}(1974)}]{Thorne74}
{Thorne}, K.~S. 1974, \apj, 191, 507

\bibitem[{{Turner} \& {Miller}(2009)}]{Turner09}
{Turner}, T.~J. \& {Miller}, L. 2009, \aapr, 17, 47

\bibitem[{{Walton} {et~al.}(2010){Walton}, {Reis}, \& {Fabian}}]{Walton10}
{Walton}, D.~J., {Reis}, R.~C., \& {Fabian}, A.~C. 2010, \mnras, 408, 601

\bibitem[{{Watanabe} {et~al.}(2004){Watanabe}, {Ohta}, {Akiyama}, \&
  {Ueda}}]{Watanabe04}
{Watanabe}, C., {Ohta}, K., {Akiyama}, M., \& {Ueda}, Y. 2004, \apj, 610, 128

\bibitem[{{{\.Z}ycki} {et~al.}(1999){{\.Z}ycki}, {Done}, \& {Smith}}]{Zycki99}
{{\.Z}ycki}, P.~T., {Done}, C., \& {Smith}, D.~A. 1999, \mnras, 309, 561

\end{thebibliography}

\newpage

    \begin{table*}
   \caption{Best-fit parameter values for the double Comptonisation model, with (second row) or without (first row) a two-phase warm absorber. $\Gamma_S$ and $kT_H$ are fixed at values $2$ and $100$ keV, respectively. Distant reflection is modelled using \textit{reflext}, whose normalisation parameter is fixed to reproduce the observed Fe $K\alpha$ line (see Section 3.3). Two Gaussian functions are included in the model, for the narrow Fe $K\beta$ and the broad Fe K emission line, with parameters fixed as in Table \ref{po}. \textit{NTHComp} normalization parameters $C$ are in units of $\textrm{keV}^{-1}\ \textrm{cm}^{-2}\ \textrm{s}^{-1}$ at $1$ keV; warm absorber column density $N_H$ is given in units of $\textrm{cm}^{-2}$, and ionisation parameter $\xi$ in units of $\textrm{erg cm s}^{-1}$. Fixed parameters are marked by (F). The last two columns give the ratio of the chi-squared value to the number of degrees of freedom and the associated probability.}              % title of Table
   \label{nthcomp_zxipcf}      % is used to refer this table in the text
   \begin{tabular}{c c c c c c c c c c c}
   \hline
   \multicolumn{11}{c}{Mrk~509}\\
   \hline
   Obs. & $kT_{S}$ (keV) & $C_{S}\ (10^{-3})$ & $\Gamma_H$ & $C_H\ (10^{-3})$ &  $N_{H,WA,1}(10^{20})$ & $log~\xi_{WA,1}$ & $N_{H,WA,2}(10^{20})$ & $log~\xi_{WA,2}$ & $\chi^2/DOF$ & prob \\
   \hline
   2+3+4 & $0.127^{+0.006}_{-0.003}$ & $0.9^{+0.2}_{-0.1}$ & $1.96^{+0.01}_{-0.01}$ & $15.8^{+0.1}_{-0.1}$ &-&-&-&-& $1077/737$ & $3\cdot10^{-15}$\\[1pt]
   2+3+4 & $0.206^{+0.005}_{-0.006}$ & $3.8^{+0.2}_{-0.2}$ & $1.92^{+0.01}_{-0.01}$ & $15.1^{+0.2}_{-0.2}$ & $8^{+2}_{-1}$ & $1.7^{+0.1}_{-0.1}$ & $7^{+9}_{-2}$ & $2.7^{+0.4}_{-0.1}$ & $800/733$ & $0.04$\\[1pt]
   \hline
   \hline
   \multicolumn{11}{c}{Mrk~841}\\
   \hline
    Obs. & $kT_{S}$ (keV)& $C_{S}\ (10^{-3})$ & $\Gamma_H$ & $C_H\ (10^{-3})$ &  $N_{H,WA,1}(10^{20})$ & $log~\xi_{WA,1}$ & $N_{H,WA,2}(10^{20})$ & $log~\xi_{WA,2}$ & $\chi^2/DOF$ & prob\\
    \hline
    1+2 & $0.137^{+0.13}_{-0.06}$ & $0.35^{+0.10}_{-0.05}$ & $1.82^{+0.01}_{-0.01}$ & $4.00^{+0.04}_{-0.08}$ &-&-&-&-& $769/707$ & $0.005$\\[1pt]
    1+2 & $0.19^{+0.01}_{-0.01}$ & $0.85^{+0.11}_{-0.11}$ & $1.80^{+0.01}_{-0.01}$ & $3.92^{+0.06}_{-0.07}$ & $14^{+4}_{-7}$ & $2.2^{+0.2}_{-0.3}$ & $<41$ & $3.3$ (F) & $705/704$ & $0.48$\\[1pt]
   \hline
   \end{tabular} 
   \end{table*}

  \begin{table*}
   \caption{Best-fit parameter values for the blurred reflection model. Distant reflection is modelled using \textit{reflext}, whose normalisation parameter is fixed to reproduce the observed narrow Fe $K\alpha$ line (see Section 3.3). A Gaussian function is included in the model, for the narrow Fe $K\beta$ emission line, with parameters fixed as in Table \ref{po}. For Mrk~509, the best-fit results including a broad Gaussian emission line are also shown. \textit{reflionx} and power-law normalization parameters $C$ are in units of $\textrm{keV}^{-1}\ \textrm{cm}^{-2}\ \textrm{s}^{-1}$ at $1$ keV; ionisation parameter $\xi$ is given in units of $\textrm{erg cm s}^{-1}$; the line normalization parameter $C_{broad}$ is the integrated number of photons $\textrm{cm}^{-2}\ \textrm{s}^{-1}$ in the line; $\sigma$ is the line width expressed in keV. The R parameter is defined as the ratio of the flux of the unabsorbed reflected component to the total unabsorbed flux (evaluated between $0.3$ and $12$ keV). The two-phase warm absorber parameters are constrained within the range evaluated with the double Comptonisation model (Table \ref{nthcomp_zxipcf}). The last two columns give the ratio of the chi-squared value to the number of degrees of freedom and the associated probability.}              % title of Table
   \label{reflionx-beta-zxipcf}      % is used to refer this table in the text
   \begin{tabular}{c c c c c c c c c c c}
   \hline
   \multicolumn{11}{c}{Mrk~509}\\
   \hline
   Obs. & $\Gamma$ & $C(10^{-3})$ & log~$\xi$ & $C_{reflionx}(10^{-6})$ & $E_{broad} (keV)$ & $C_{broad}(10^{-5})$ & $\sigma_{broad} (keV)$ & R & $\chi^2/DOF$ & prob\\
   \hline
  2+3+4 & $2.041^{+0.004}_{-0.004}$ & $15.9^{+0.1}_{-0.1}$ & $1.54^{+0.14}_{-0.06}$ & $11.5^{+2.2}_{-2.9}$ & - & - & - & $0.16^{+0.03}_{-0.04}$ & $915/733$ & $5\cdot10^{-6}$\\[1pt]
   2+3+4 & $2.039^{+0.007}_{-0.005}$ & $15.9^{+0.1}_{-0.1}$ & $1.73^{+0.03}_{-0.02}$ & $6.1^{+0.7}_{-1.1}$ & $6.9^{+0.1}_{-0.1}$ & $10^{+1}_{-1}$ & $0.8^{+0.3}_{-0.2}$ & $0.15^{+0.01}_{-0.02}$ & $756/730$ & $0.25$ \\[1pt]
   \hline
   \hline
   \multicolumn{11}{c}{Mrk~841}\\
   \hline
    Obs. & $\Gamma$ & $C(10^{-3})$ & log~$\xi$ & $C_{reflionx}(10^{-6})$ & $E_{broad} (keV)$ & $C_{broad}(10^{-5})$ & $\sigma_{broad} (keV)$ & R & $\chi^2/DOF$ & prob\\
    \hline
   1+2 & $1.89^{+0.02}_{-0.01}$ & $3.9^{+0.1}_{-0.1}$ & $1.80^{+0.09}_{-0.05}$ & $1.9^{+0.04}_{-0.05}$ & - & - & - & $0.18^{+0.04}_{-0.05}$ & $730/704$ & $0.24$\\[1pt]
   \hline
   \end{tabular} 
   \end{table*}

   \begin{table*}
   \caption{Best-fit parameter values for the smeared absorption model. Distant reflection is modelled using \textit{reflext}, whose normalisation parameter is fixed to reproduce the observed Fe $K\alpha$ line (see Section 3.3). For both sources, two Gaussian functions are included in the model, for the narrow Fe $K\beta$ and the broad Fe K emission line, with parameters fixed as in Table \ref{po}. Power-law normalization parameter $C$ is in units of $\textrm{keV}^{-1}\ \textrm{cm}^{-2}\ \textrm{s}^{-1}$ at $1$ keV; wind column density $N_H$ is given in units of $\textrm{cm}^{-2}$, and the ionisation parameter $\xi$ in units of $\textrm{erg cm s}^{-1}$. The two-phase warm absorber parameters are constrained within the range evaluated with the double Comptonisation model (Table \ref{nthcomp_zxipcf}).  The last two columns give the ratio of the chi-squared value to the number of degrees of freedom and the associated probability.}              % title of Table
   \label{swind1_zxipcf}      % is used to refer this table in the text
   \begin{tabular}{c c c c c c c c}
   \hline
   \multicolumn{8}{c}{Mrk~509}\\
   \hline
   Obs. & $\Gamma$ & $C(10^{-3})$ & $N_H (10^{22})$ & $log\ \xi$ & $\sigma$ & $\chi^2/DOF$ & prob \\
   \hline
   2+3+4 & $2.06^{+0.01}_{-0.01}$ & $23.8^{+0.3}_{-0.5}$ & $18^{+2}_{-1}$ & $3.29^{+0.03}_{-0.02}$ & $>0.45$ & $748/732$ & $0.33$\\[1pt]
   \hline
   \hline
   \multicolumn{8}{c}{Mrk~841}\\
   \hline
   Obs. & $\Gamma$ & $C(10^{-3})$ & $N_H (10^{22})$ & $log\ \xi$ & $\sigma$ &  $\chi^2/DOF$ & prob\\
   \hline
   1+2 & $1.97^{+0.02}_{-0.02}$ & $6.25^{+0.07}_{-0.12}$ & $12^{+3}_{-3}$ & $3.10^{+0.10}_{-0.09}$ & $>0.46$ & $732/703$ & $0.21$\\[1pt]
   \hline
   \end{tabular} 
   \end{table*}
   
   \setcounter{figure}{6}

      \begin{sidewaysfigure*}
	 \centering
   \resizebox{\hsize}{!}{\includegraphics[scale=1]{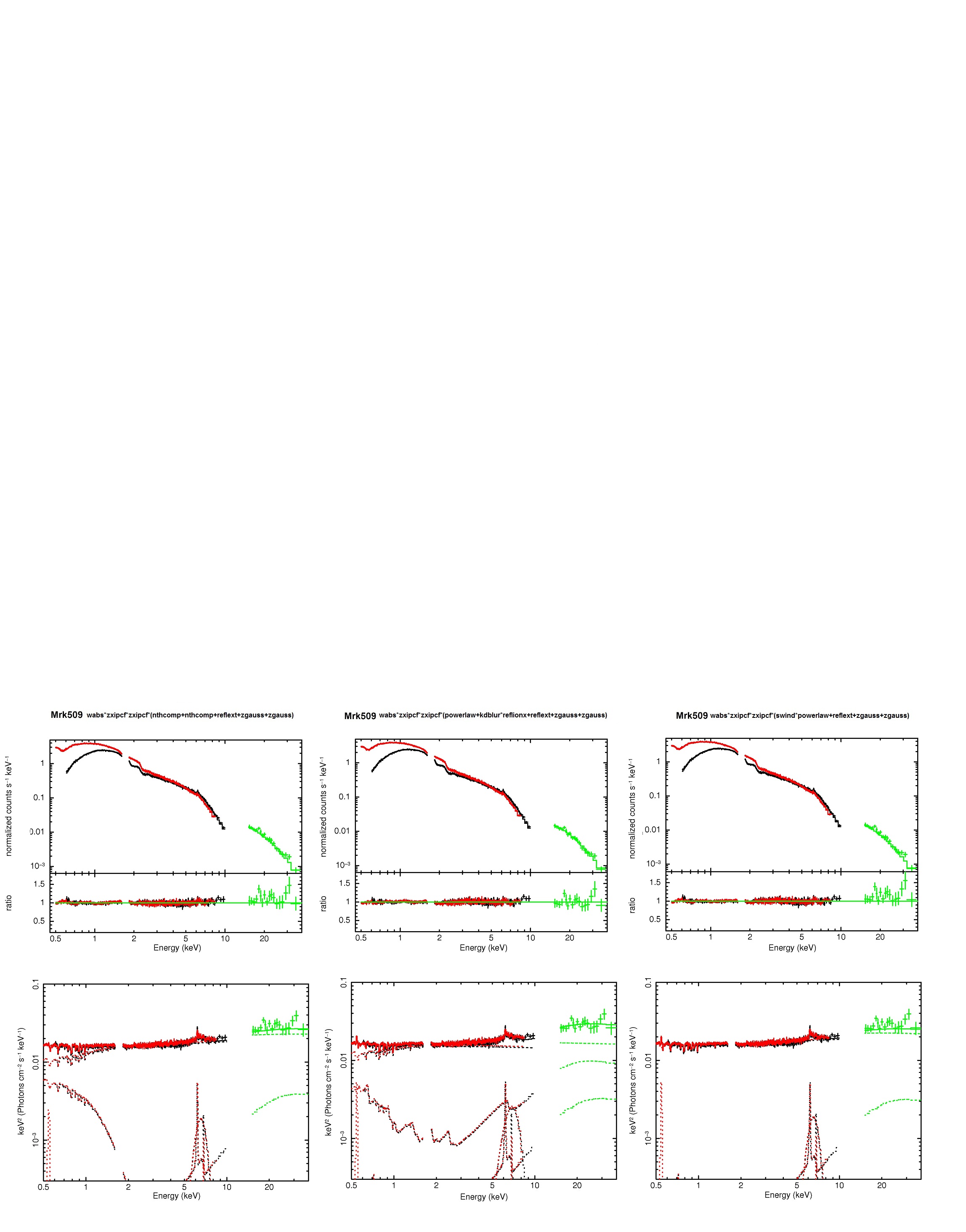}}
     \caption{Top: Mrk~509 data (sum of \textit{Suzaku} observations) and, in the subplot, data-to-model ratio. Bottom: unfolded data. Colour code for the different \textit{Suzaku} instruments: \textit{XIS/FI}-black, \textit{XIS1}-red, and \textit{HXD/PIN}-green. Three different models are shown, from left to right: double Comptonisation, blurred reflection and smeared absorption. Solid lines correspond to the best-fit model, and thin dashed lines to the model different components. For the three models a broad emission-line component is required. The energy on the abscissa is in the observer frame. The data were rebinned for clarity purposes.}
     \label{Mrk509_ldata}
   \end{sidewaysfigure*}
   
  \begin{sidewaysfigure*}
	 \centering
   \resizebox{\hsize}{!}{\includegraphics[scale=1]{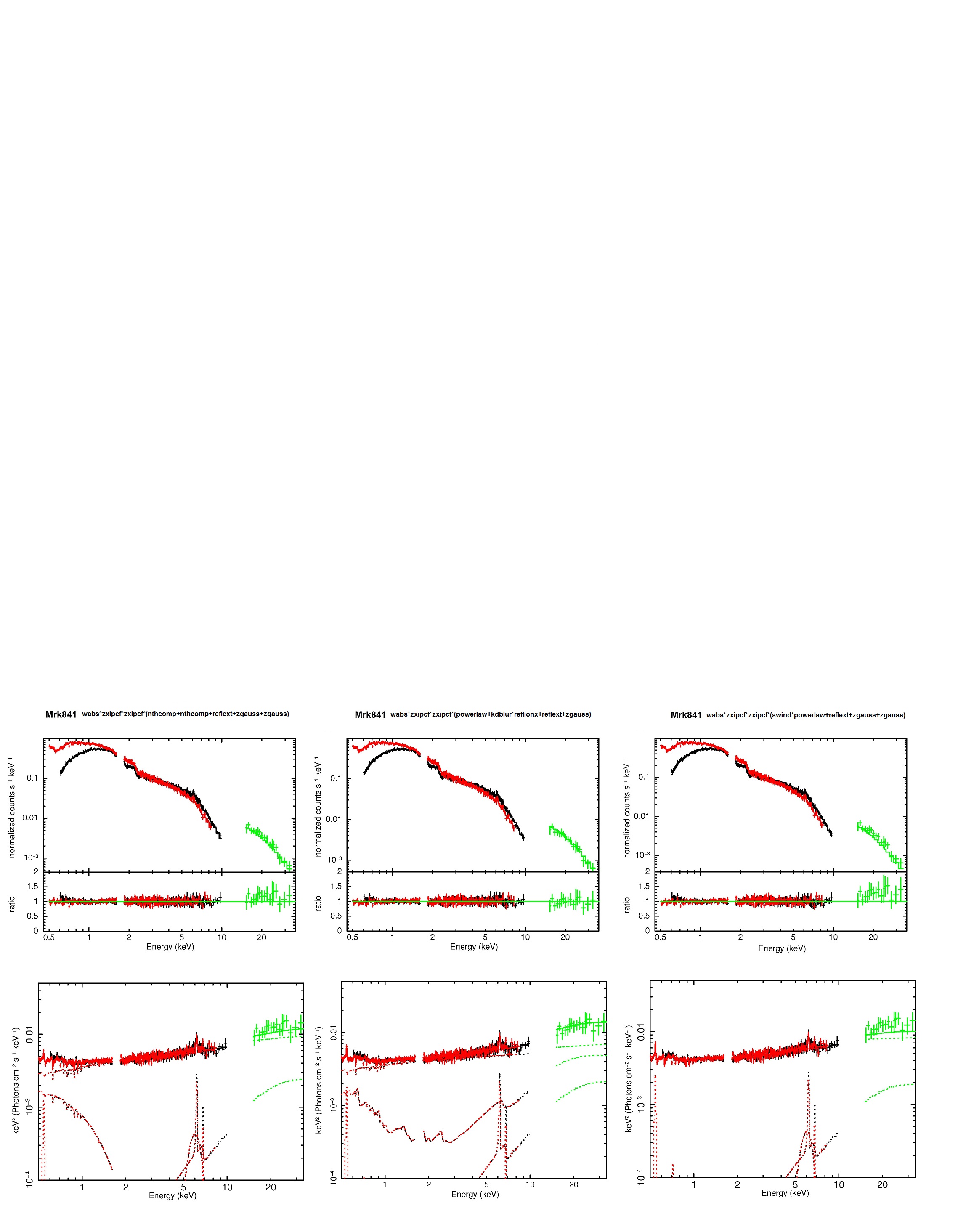}}
     \caption{Top: Mrk~841 data (sum of \textit{Suzak} observations) and, in the subplot, data-to-model ratio. Bottom: unfolded data. Colour code for the different \textit{Suzaku} instruments: \textit{XIS/FI}-black, \textit{XIS1}-red, and \textit{HXD/PIN}-green. Three different models are shown, from left to right: double Comptonisation, blurred reflection and smeared absorption. Solid lines correspond to the best-fit model, and thin dashed lines to the model different components. A broad emission-line component is added for the double Comptonisation and the smeared absorption model only. The energy on the abscissa is in the observer frame. The data were rebinned for clarity purposes.}
     \label{Mrk841_ldata}
   \end{sidewaysfigure*}   
  
   \end{document}